**Reports on Progress in Physics**

**Review Title:**

**Two-photon imaging and analysis of neural network dynamics**

**Authors:** Henry Lütcke & Fritjof Helmchen, Brain Research Institute, University of Zurich

**Abstract**

The glow of a starry night sky, the smell of a freshly brewed cup of coffee or the sound of ocean waves breaking on the beach are representations of the physical world that have been created by the dynamic interactions of thousands of neurons in our brains. How the brain mediates perceptions, creates thoughts, stores memories and initiates actions remains one of the most profound puzzles in biology, if not all of science. A key to a mechanistic understanding of how the nervous system works is the ability to measure and analyze the dynamics of neuronal networks in the living organism in the context of sensory stimulation and behaviour. Dynamic brain properties have been fairly well characterized on the microscopic level of individual neurons and on the macroscopic level of whole brain areas largely with the help of various electrophysiological techniques. However, our understanding of the mesoscopic level comprising local populations of hundreds to thousands of neurons (so called 'microcircuits') remains comparably poor. In large parts, this has been due to the technical difficulties involved in recording from large networks of neurons with single-cell spatial resolution and near-millisecond temporal resolution in the brain of living animals. In recent years, two-photon microscopy has emerged as a technique which meets many of these requirements and thus has become the method of choice for the interrogation of local neural circuits. Here, we review the state-of-research in the field of two-photon imaging of neuronal populations, covering the topics of microscope technology, suitable fluorescent indicator dyes, staining techniques, and in particular analysis techniques for extracting relevant information from the fluorescence data. We expect that functional analysis of neural networks using two-photon imaging will help to decipher fundamental operational principles of neural microcircuits.

**Contents**

1. **Introduction**
2. **Imaging neural networks in vivo**
   - 2.1. Two-photon microscopy of neuronal populations
   - 2.2. Small molecule and protein indicators of neural activity
   - 2.3. Delivery of activity indicators to brain tissue

## 1. Introduction

During the past century the prevailing technique for studying neuronal activity in living brains ('in vivo') has been the electrical recording of neural spikes, the extracellular signature of action potentials (APs) generated by neurons. This approach has been employed successfully for more than half a century on the microscopic level by extracellular recordings of individual neurons (Jung et al., 1957; Mountcastle et al., 1957; Hubel and Wiesel, 1959). Spikes have been recorded from individual neurons (single "units"), from a few separable units, or from groups of inseparable units near the electrode tip ("multi-unit" recordings). Even though simultaneous extracellular recordings from tens to hundreds of neurons are possible (Buzsaki, 2004), these measurements typically are obtained from spatially dispersed sets of neurons. Other limitations of extracellular recordings are poorly defined identity of recorded cells (for example with regard to neuronal subtype or laminar position) and blindness for silent neurons. Another electrophysiological technique that has been applied in vivo is intracellular recording using either sharp glass electrodes or the patch-clamp technique (Chorev et al., 2009). With intracellular recording the subthreshold membrane potential dynamics can be resolved, in particular synaptically-induced potentials. However, this technique is limited to single neurons (or pairs of neurons at most; see Okun and Lampl, 2008; Poulet and Petersen, 2008). Thus, despite the power of electrophysiological methods they so far are unable to resolve the activity state of a population of nearby neurons (a "local microcircuit") in the living brain. In order to gain further insights into the principles of signal flow through a neural circuit under conditions of relevant operation (e.g. during a specific behaviour), comprehensive in vivo measurements of microcircuit dynamics with the capability to resolve the temporal evolution of the neural network state are required.

      End of the 20[th] century two-photon excited fluorescence laser scanning microscopy (Denk et al., 1990) - short 'two-photon microscopy' - has emerged as a novel optical imaging techniques that now enables neuroscientists to measure neural network dynamics in live animals more comprehensively. Two-photon microscopy has revolutionized in vivo imaging by enabling fluorescence imaging several hundreds of micrometer deep in intact tissue and with spatial resolution sufficient to discern individual neurons, dendrites, and synapses in the neocortex of living rats and mice (for reviews see Denk and Svoboda, 1997; Helmchen and Waters, 2002; Helmchen and Denk, 2005; Svoboda and Yasuda, 2006; Kerr and Denk, 2008). No other microscope technique - including

confocal microscopy - achieves comparable spatial resolution and signal contrast from inside scattering tissue, therefore our review is fully focused on the application of two-photon microscopy. Two-photon microscopy meanwhile has developed into the key technology for studying dynamic processes on the cellular and subcellular scale in living tissue, not only in the brain but also in skin, kidney, lymph nodes etc. In neuroscientific applications, neurons and their dendrites can be now routinely visualized down to 0.5 mm depth below the surface of animal brains and, in special cases, down to nearly 1 mm. Two-photon microscopy has proven particularly powerful in combination with fluorescent reporters of neuronal activity. Such indicators (either small organic molecules or proteins) change their fluorescence properties as a function of either membrane potential or intracellular calcium concentration, both reporting the activity state of neurons. A comprehensive functional analysis of neural networks will require simultaneous cell-type specific measurements of many individual neurons in the context of their spatial locations within the microcircuit. With the help of new labelling approaches, especially for calcium indicators, it has become possible to stain entire local populations of neurons, for example in the neocortex, and thereby enable imaging of neural network dynamics (reviewed in Garaschuk et al., 2006; Grewe and Helmchen, 2009). Since many neocortical areas operate in a sparse firing regime (Kerr et al., 2007; Hromadka et al., 2008; Wolfe et al., 2010) monitoring of single action potentials in neurons, preferentially with near-millisecond temporal resolution, is essential. Lastly, it is desirable to achieve such measurements in the intact or even awake animal in order to be physiologically relevant. In the following, we will summarize recent advances in this field, focusing on methods for two-photon imaging of neuronal population activity and their recent applications to analyze network dynamics. At the end of the review we will discuss future prospects.

## 2. Imaging neural networks in vivo

### 2.1. Two-photon microscopy of neuronal populations

What is special about two-photon microscopy that it has enabled in vivo network imaging? The key advantage of two-photon microscopy is its lower sensitivity to light scattering compared to other fluorescence techniques, e.g., confocal microscopy (Denk et al., 2005; Helmchen and Denk, 2005; Svoboda and Yasuda, 2006). Light scattering usually destroys optical resolution and contrast in biological tissue. The trick in two-photon microscopy is the special form of fluorescence excitation, "two-photon excitation", which is a nonlinear process that requires quasi-simultaneous absorption of two near-infrared photons (typically a Titanium:sapphire laser with 720-1050 nm tuneable output is used). The probability $n_a$ for such a two-photon absorption event is given by (Denk et al., 1990)

$$n_a \propto \frac{\delta_2 \; P_{avg}^2}{\tau_p \, f_p^2} \; \frac{NA^4}{(hc\lambda)^2} \tag{1}$$

where $P_{avg}$ is the average laser power, $\delta_2$ is the two-photon excitation cross section (measured in units of Göppert-Mayer (GM); in honour of Maria Göppert-Mayer for predicting the two-photon absorption process Göppert-Mayer, 1931), $\tau_p$ is the laser pulse width, $f_p$ the pulse repetition rate, $\lambda$ the excitation wavelength, and NA the numerical aperture of the objective. Due to the low probability of this process, a sufficient rate of two-photon excitation for fluorescence imaging requires high photon density, which is usually achieved by using a pulsed laser source and spatially focusing the laser beam through a high-NA microscope objective. The factor $(\tau_p f_p)^{-1}$ is often referred to as the "two-photon advantage", describing the gain in fluorescence yield, which is achieved by using a pulsed instead of a continuous-wave laser source. For Titanium:sapphire lasers, which provide laser pulses of about 100 fs width with a repetition rate on the order of 100 MHz, the two-photon advantage is on the order of $10^5$. Besides the longer wavelength, which makes the excitation light less susceptible to scattering, the key advantage of two-photon excitation is its quadratic dependence on the average laser power, due to which fluorescence generation is confined to the micron-sized focal spot and excitation in out-of-focus sections is abolished. Most simply, the laser focus is scanned in a raster-like fashion across the *x-y* plane (orthogonal to the optical axis which is referred to as *z*-axis) to generate an image (figure 1). Intrinsic optical sectioning is a chief property of two-photon microscopy because it implies that the exact place of origin of fluorescence photons is well-defined at each point in time. Consequently, photons multiply scattered on their way out of the tissue can still be correctly assigned to the respective volume element. This notion implies that even in strongly scattering tissue (such as the brain) all photons carry useful signal and the detection strategy becomes very simple: 'just collect as many photons as you can' (Denk et al., 2005). For further details on the physical principles and technical aspects of two-photon microscopy we refer to several comprehensive reviews (So et al., 2000; Diaspro, 2002; Zipfel et al., 2003; Denk et al., 2005; Helmchen and Denk, 2005; Carriles et al., 2009).

For about 8 years, it has been possible to stain entire populations of neurons in local brain regions in living animals with activity-dependent fluorescent indicators. The key methods and the various approaches that have been developed are summarized further below. One example is shown in figure 1, where a population of neocortical neurons has been stained with a fluorescent calcium indicator using a 'multi-cell bolus loading' technique (Stosiek et al., 2003). With the ability to achieve comprehensive in vivo staining of hundreds of neurons with functional indicators, it has become a challenge to optimize two-photon microscopy technology for the readout of network activity (for recent reviews see Kerr and Denk, 2008; Grewe and Helmchen, 2009). Here we provide a brief overview of recent advances. For our discussion two aspects are particularly relevant: imaging speed and sample size, i.e., number of neurons imaged. Obviously, it is desirable to image large populations at high speed in order to reveal neuronal spike trains across the population with good temporal precision, at best with millisecond resolution. However, because the number of photons is limited in such optical recordings due to the small cell compartments investigated and low fluorescence yields, the signal-to-noise ratio (SNR) of the fluorescence measurement becomes a key factor in determining, which combinations of speed and sample size are feasible. Even when neglecting all other noise sources, fluorescence measurements are fundamentally limited by shot noise, for which the standard deviation of a fluorescence measurement scales with $\sqrt{N_{ph}}$, where $N_{ph}$ denotes the average number of detected photons. Thus, sufficient numbers of fluorescence photons need to be collected (by efficient detection optics and by dwelling long enough on each cell) in order to reduce noise to acceptable levels. What noise level is still acceptable depends on the specific information that one aims to extract from the fluorescence recordings, for example whether one aims to obtain a coarse spatial map of neuronal activation or whether one needs to reveal spike trains with single action potential (AP) resolution and high temporal precision. Here we define the SNR as the ratio of the fluorescence change evoked by an elementary neuronal excitation event, e.g. the action potential, and the standard deviation of the baseline fluorescence trace. For a given SNR, the sample size $N_{cells}$ and the acquisition rate $f_{acq}$ follow an inverse relationship: either activity is recorded from a small group of neurons at highest speed, or a large population is sampled at low to moderate speed. These relationships are summarized in the simple equation

$$N_{cells} \ f_{acq} \ SNR = const. \tag{2}$$

highlighting the trade-off that has to be made between speed and network size. Clearly, the exact relationship (in particular the value of the constant) depends on multiple factors such as the fluorescence indicator used, the spatial distribution of neurons, and the specific laser scanning mode.

The default acquisition mode for two-photon imaging is raster-like frame scanning, which, however, is not optimized for fast network measurements. Depending on the microscope objective and the zoom factor, a field-of-view of 100-800 μm side length permits imaging of tens to hundreds of neurons, albeit typically with relatively slow frame rates (1-10 Hz for image resolutions between 128x128 and 512x512 pixels). Recent variants of two-photon microscopy have pushed the limits for both sample size and speed further, employing novel scanning schemes that are tuned to efficiently collect data from neuronal populations. In general, these special methods avoid – as much as possible – the waste of precious scan time on irrelevant regions (e.g. background) and attempt to maximize fluorescence collection from the cells of interest. One simple concept is to employ user-defined line scans, where a scan trajectory of the laser focus is freely chosen or calculated based on a set of pre-selected neurons (figure 1). The dwell time on the cells can be further optimized by accelerating scanning on the path segments between cells and slowing it down on each cell (Lillis et al., 2008). With this approach calcium signals have been measured in populations of several tens of neurons at 30-300 Hz acquisition rate (Göbel and Helmchen, 2007b; Lillis et al., 2008; Rothschild et al., 2010; Valmianski et al., 2010), usually requiring low-pass filtering to reach a sufficient SNR. The main limitation of this approach arises from the inertia of the standard galvanometric scan mirrors, restricting their frequency response characteristics to below a few kilohertz.

A second option for gaining speed is to use a resonant galvanometric mirror for the fast scan axis (Leybaert et al., 2005; Rochefort et al., 2009). Typical resonance frequencies are 1-12 kHz and therefore allow frame rates of about 60 Hz (for 128-256 lines). Low-noise calcium measurements at video rate from populations of about 20–30 cells have been reported (Rochefort et al., 2009). The rapid scanning may provide an advantage because the very brief dwell time of the laser focus on each cell (about 5-10 μs per crossing) possibly reduces fluorophore exhaustion by preventing molecular transition into non-fluorescent 'dark' states (Donnert et al., 2007). On the other hand, drawbacks of resonant scanning are its limited flexibility and the still significant amount of time wasted on background regions.

Further optimization of the usage of scan time can be achieved with a non-mechanical laser scanning approach. This technique is based on acousto-optical deflectors (AODs), which diffract a laser beam at a sound wave that is induced in a glass crystal with a piezoelectric transducer (for review see Saggau, 2006). Here, the direction of laser beam diffraction can be changed arbitrarily within the scan range in a few microseconds (figure 1). Two crossed AODs therefore enable very fast random-access scanning in the *x-y* plane (Iyer et al., 2006; Otsu et al., 2008). The implementation of AOD-based laser scanning for two-photon microscopy has been somewhat slowed by the necessity of special optical arrangements to compensate for laser beam distortions and laser pulse broadening induced by optical dispersion (Zeng et al., 2006; Kremer et al., 2008). Recently, we have applied an AOD-based two-photon microscope for high-speed measurements of neuronal network activity in vivo. We devised a special random-access pattern scanning mode, in which each cell within a preselected group of neurons is sampled with a small pattern of scan points (total dwell time about 50 μs), and achieved measurements from about 30-80 neurons at 200-500 Hz sampling rate (Grewe et al., 2010). Most notably, the SNR of unfiltered data was sufficiently high (2-4) to deduce the time of occurrence of individual spikes with about 5-15 ms precision. Another recent report using similar techniques in brain slices came to similar conclusions (Ranganathan and Koester, 2010). The AOD-scanning approach thus is highly promising for optically resolving whole sets of spike trains within local neuronal populations, reaching a temporal resolution almost comparable to electrical recordings.

Finally, various other scanning approaches are still being explored. One direction is the use of multi-focal approaches, where several laser foci are applied in parallel (Kurtz et al., 2006; Niesner et al., 2007). Using holographic techniques it is even possible to shape the excitation pattern, e.g. to excite a specific subset of neuronal cell bodies (Nikolenko et al., 2008). While these multi-foci approaches may be well suited for experiments in brain slices, their fundamental drawback is the sensitivity to scattering when imaging deep in tissue, because cross-talk between multiple excitation beams is hard to avoid. Another major direction of current developments is the extension of fast population measurements to the third dimension in space (figure 1). Most simply, the free line scan approach has been extended to 3D in order to enable scanning of large cell populations distributed in a local volume, e.g., in cortical layer 2/3 (Göbel et al., 2007). Scanning along the *z*-axis was achieved with a piezoelectrically driven focusing device moving the microscope objective and calcium signals were measured from up to 500 cells at an effective sampling rate per cell of 10 Hz. Various novel approaches for fast *z*-scanning - including remote mechanical *z*-scanning (Botcherby et al., 2007), temporal focusing (Durst et al., 2006), and variable focus lenses (Olivier et al., 2009) - are currently being explored and will likely make large-scale volumetric network measurements more practicable in the near future. For example, by splitting the laser beam in several sub-beams and reading out the generated fluorescence in a time-multiplexed fashion, simultaneous high-speed in vivo recordings from multiple focal planes have recently been achieved (Cheng et al., 2011). Moreover, the non-mechanical AOD-based scanning approach in principle can be also be extended to 3D (Duemani Reddy et al., 2008; Kirkby et al., 2010).

In summary, the last years have seen rapid progress in two-photon microscopy techniques, in particular with regard to special laser scanning methods for reading out neuronal network dynamics. All of these methods rely on the use of sensitive activity-dependent fluorescent indicators, which we will describe in the next section.

## 2.2. Small molecule and protein indicators of neural activity

Part of the power of microscopy techniques in biology derives from the wide range of fluorophores available for staining biological tissue. Because emission light can be collected at multiple wavelengths, different types of fluorophores can be employed to obtain anatomical and functional information simultaneously. Two-photon microscopy, in particular, benefits from the availability of multiple fluorophores as the two-photon absorption spectrum of most dyes is relatively broad and different dyes can thus be efficiently excited at a single wavelength.

Historically, fluorescence microscopy in neuroscience has relied on a wide range of small, organic dyes for tissue labelling, which allow visualization of structural and dynamic aspects of the nervous system. With the cloning of green fluorescent protein (GFP) from the jellyfish *Aequorea victoria* as well as related proteins from corals, genetically encoded sensors of cell activity have become feasible (Miyawaki et al., 1997). Compared to organic dyes, protein sensors possess several advantages that make them particularly attractive for neuroscientists. Most importantly, since proteins are encoded by DNA, they can be directly expressed in the tissue of interest, ideally for long time periods. If sensors are expressed under the control of different promoters, neuron or astrocyte specific expression can be achieved (Tian et al., 2009; Lütcke et al., 2010). Moreover, by including localization signals in the DNA sequence, genetically-encoded probes can even be localized to specific sub-cellular organelles, such as synapses (Dreosti et al., 2009) or the plasma membrane (Shigetomi et al., 2010). In combination with conditional gene expression strategies, such as the Cre-Lox system (see below), expression of fluorescent proteins may be even more tightly regulated, for example to specific sub-types of neurons or developmental stages.

*Optical monitoring of neural activity: voltage sensors.* Neurons integrate synaptic input and communicate with each other by the propagation of electrical signals along their processes. Measuring membrane depolarization therefore constitutes the most direct measure of neural activity.

Consequently, the first optical recordings of neural network activity were performed using the voltage-sensitive dye (VSD) merocyanine (Salzberg et al., 1977). Subsequently, a number of different VSDs have been designed that stain nervous tissue more efficiently and report membrane potential changes with higher sensitivity, thus allowing multi-site membrane potential measurements in the mammalian brain in vivo (Arieli et al., 1996; Petersen et al., 2003), albeit without cellular resolution. VSDs are typically lipophilic dyes that incorporate into the cell membrane where they change fluorescence properties (e.g. brightness) upon membrane potential changes. Voltage-sensitive dyes allow imaging with exquisite temporal resolution in the millisecond range and report both subthreshold and suprathreshold membrane changes, including inhibitory synaptic potentials. VSDs are thus uniquely suitable for imaging the temporal spread of activation across large parts of the brain (see for example Ferezou et al., 2007).

While organic VSDs have been used to image neural dynamics for more than three decades (Grinvald and Hildesheim, 2004), recent efforts have focused on the development of voltage-sensitive fluorescent proteins to achieve more efficient, cell-type specific staining (for review see Baker et al., 2008). In a first attempt, several groups fused fluorescent proteins (GFP, CFP / YFP) with the voltage-sensing domains of different voltage-gated ion channels (Siegel and Isacoff, 1997; Sakai et al., 2001; Ataka and Pieribone, 2002). These first generation sensors optically reported changes in membrane potential, however, their widespread use was precluded by poor localization of the sensor to the plasma membrane resulting in significant background fluorescence. More recently, voltage-sensing domains from proteins other than ion channels were fused with yellow and cyan fluorescent proteins to generate novel ratiometric voltage-sensitive fluorescent proteins (VSFPs, Dimitrov et al., 2007). The newest VSFPs (VSFP2.3, VSFP2.42) are characterized by efficient membrane targeting as well as an improved dynamic range optimized for in vivo voltage changes and were shown to report cortical responses to single-whisker stimulation in living mice (Akemann et al., 2010). With the development of even better VSFPs we expect to soon see widespread use of these sensors for cell-type specific monitoring of neural circuit dynamics with high temporal resolution. Although organic and protein voltage sensors possess excellent temporal properties, they still suffer from poor signal-to-noise ratio compared to calcium sensors (see below), thus often requiring extensive spatial averaging. VSD imaging in vivo still lacks single-cell spatial resolution and typically cannot distinguish between signals originating from different depths of the cortex (but see Kuhn et al., 2008). Finally, it should be noted that application of VSDs may result in unwanted side effects such as tissue damage or even alterations of cellular physiology, as has recently been demonstrated (Mennerick et al., 2010).

*Optical monitoring of neural activity: calcium sensors.* Neuronal membranes contain voltage-gated calcium channels that open when a cell elicits an action potential, leading to a brief influx of calcium ions and a transient increase in the intracellular calcium concentration $[Ca^{2+}]_i$ (Helmchen et al., 1996). These AP-evoked $[Ca^{2+}]_i$ transients are reported by fluorescent calcium indicators as they change their fluorescence properties upon $Ca^{2+}$ binding (figure 1). For most neurons, $[Ca^{2+}]_i$ increases in the neuronal cell body are tightly coupled to AP firing, therefore indicator fluorescence changes are a good measure of supra-threshold activity. Due to the large fractional fluorescence changes attainable, calcium indicators at present are the most sensitive reporters of neural activity, with some of them resolving single AP-evoked $[Ca^{2+}]_i$ transients in networks of identified neurons in vivo (Kerr et al., 2005). Following the rapid (within milliseconds) $[Ca^{2+}]_i$ increase after an AP, $[Ca^{2+}]_i$ decays back to resting level on a longer time scale of hundreds of milliseconds as calcium ions are buffered within the cytosol and removed via various extrusion mechanisms. Most simply, the stereotype event of a single AP-evoked $[Ca^{2+}]_i$ transient is formally approximated by an immediate step increase of amplitude A with a subsequent exponential decrease with decay constant $\tau$:

$$\Delta\left[Ca^{2+}\right] = A\, e^{-(t-t_0)/\tau} \qquad \text{for t} \geq t_0 \qquad (3)$$

where $t_0$ is the time point when the AP occurs. Amplitude and $\tau$ depend on various factors, including the concentration and properties of the calcium indicator itself (for a detailed treatment see Helmchen, 2011; Helmchen and Tank, 2011). With the advent of high-speed imaging systems, the fluorescence response to a single AP may be more accurately described by a fast exponential onset ($\tau_{on} \approx 10 - 20$ ms for synthetic indicators (Grewe et al., 2010); and $\tau_{on} \approx 100$ ms for genetically-encoded probes (Tian et al., 2009)) as well as a comparatively slow single- or double-exponential decay ($\tau_{fast}$ and $\tau_{slow}$), again depending on the experimental conditions:

$$\Delta\left[Ca^{2+}\right] = (1 - e^{-(t-t_0)/\tau_{on}})(A_{fast}\, e^{-(t-t_0)/\tau_{fast}} + A_{slow}\, e^{-(t-t_0)/\tau_{slow}}) \qquad \text{for t} \geq t_0 \qquad (4)$$

Here, $A_{fast}$ and $A_{slow}$ refer to the relative contributions of the fast and the slow components, respectively. For the synthetic calcium indicator Oregon Green BAPTA-1 (OGB-1) typical relative fluorescence changes evoked by a single AP in neuronal somata are about 7-8% (Kerr et al., 2005; Grewe et al., 2010). Furthermore, with the new high-speed two-photon calcium imaging methods it is now possible to infer the timing of APs from the onsets of the fluorescence traces with near-millisecond precision (Grewe et al., 2010).

Since the early 1980s, a large number of fluorescent calcium indicators have been designed based on the calcium chelator BAPTA, showing large calcium-induced fluorescence changes, high specificity for $Ca^{2+}$ as opposed to other ions such as $Mg^{2+}$ as well as rapid kinetics (for review seeTsien, 1989). A variety of dyes with different excitation and emission wavelengths are available. Most organic calcium indicators are also available in membrane-permeable acetoxymethyl (AM) ester form to allow bulk loading of large numbers of cells in the intact brain (see below). Among the dyes that have been used for in vivo imaging of neural network dynamics are Oregon Green BAPTA-1 (OGB-1) AM (figure 1) (Kerr et al., 2005), Fura-2 AM (Sohya et al., 2007), Fluo-4 AM (Sato et al., 2007), and various others (Stosiek et al., 2003). Similar to small-molecule VSDs, organic calcium indicators suffer from a number of shortcomings. First, bulk labelling with AM-ester dyes in most cases indiscriminately stains all cell types, including excitatory and inhibitory neurons as well as astrocytes, thus impeding cell-type specific measurements. While neurons and astrocytes can be distinguished relatively easily using additional stains (Nimmerjahn et al., 2004), distinction between different neuronal subtypes, such as pyramidal and interneurons, requires additional experimental sophistication (see below). Second, AM-ester dyes gradually leak from the cytosol into the extracellular space and into intracellular compartments over the duration of several hours. Relabeling of the same cell populations has proven challenging (but see Andermann et al., 2010) and imaging experiments are usually limited in duration to several hours. Finally, in vivo bulk loading of AM-dyes leads to substantial fluorescence in the neuropil, staining dendritic and axonal processes in the area of interest. Care must be taken to clearly separate diffuse calcium signals in this neuropil background stain from the calcium signals in individual neurons (Kerr et al., 2005; Göbel and Helmchen, 2007a; Kerr and Denk, 2008). Due to the relatively high background fluorescence, visualization of individual cellular processes such as dendrites through MCBL occurs only under special conditions (Kerr et al., 2005).

Recent progress in the development of genetically encoded calcium indicators (GECIs) is beginning to address some of these concerns (for reviews see Miyawaki, 2003; Hires et al., 2008; Mank and Griesbeck, 2008). In principle, GECIs and VSFPs are designed according to similar blueprints. Instead of a voltage sensor, GECIs contain a $Ca^{2+}$-binding protein domain, such as found in calmodulin, which undergoes a conformational change between $Ca^{2+}$-bound and $Ca^{2+}$-free form (figure 2). Frequently, the conformational change is enhanced by fusion of a conformational actuator (such as M13) which preferentially binds to the $Ca^{2+}$-bound form of calmodulin. GECIs are classified in two groups based on the number of fluorescent proteins (FPs) fused to the $Ca^{2+}$-sensitive protein part. For single-FP GECIs the $Ca^{2+}$-sensing protein part is linked to a single FP, typically an enhanced (and possibly circularly permuted) version of GFP (eGFP). Fluorescence properties (brightness or emission wavelength) of the eGFP are modulated by changes in the reporter configuration thus imposing calcium sensitivity upon the protein (prominent examples are the G-CaMPs). The second class of GECIs contains two different fluorescent proteins that can form a fluorescence resonance energy transfer (FRET) pair. Typically a cyan FP (CFP) variant acts as FRET donor while a yellow FP (YFP) acts as FRET acceptor (figure 2). The calcium-induced conformational change in the sensor protein brings the two FPs in closer proximity and thereby increases the CFP to YFP FRET efficiency (which depends steeply on the distance between fluorophores). In practice, a decrease in CFP and an increase in YFP fluorescence intensity are observed, such that the ratio between CFP and YFP provides a sensitive read-out of $[Ca^{2+}]_i$. Ratiometric GECIs have a number of advantages for in vivo imaging as will be discussed below.

Since the first report of a GECI ('cameleon' by Miyawaki et al., 1997), numerous variants of 1-FP and 2-FP GECIs have been constructed (see Mank and Griesbeck, 2008 for a comprehensive overview). Early GECI versions have been designed based on calmodulin, which is itself an abundant $Ca^{2+}$-binding protein in neurons. This has prompted concerns that expression of calmodulin-based GECIs may interfere substantially with neuronal calcium signalling and buffering. In response,

Griesbeck and colleagues (Heim and Griesbeck, 2004) introduced a series of ratiometric GECIs based on calcium sensing protein Troponin C, which is expressed in muscles but not neurons (Heim and Griesbeck, 2004). Early GECIs furthermore suffered from poor sensitivity to $[Ca^{2+}]_i$ and, unlike organic dyes, were not able to detect low numbers of APs in vivo (Mao et al., 2008). In recent years, a number of improved GECIs have been constructed that represent significant advances on earlier versions, especially with respect to calcium sensitivity. Wallace et al. reported optical detection of single APs in the mouse barrel cortex in vivo with the ratiometric indicator d3cpV (Wallace et al., 2008). Similarly, the 1-FP reporter G-Camp3 showed high AP-sensitivity as well as large dynamic range and, crucially, could be used to monitor spiking activity in awake animals over months (Tian et al., 2009). Recently, we have shown that virus-mediated delivery of the GECI Yellow Cameleon 3.60 (YC3.60; Nagai et al., 2004) specifically to neurons (using a synapsin promoter) enables in vivo measurement of neuronal activity with high fidelity and across multiple spatial scales (Lütcke et al., 2010). By combining two-photon calcium imaging with cell-attached recordings, we could show that YC3.60 can resolve both spontaneously occurring and sensory-evoked single APs in vivo and reliably reports bursts of APs with negligible saturation (figure 2). Furthermore, $Ca^{2+}$ transients could not only be detected in populations of neurons but even in individual apical dendrites. Finally, in combination with fibre-optic recording techniques, YC3.60 enabled measurement of complex $Ca^{2+}$ dynamics in awake, freely moving mice. Together, our results and those from other groups demonstrate that GECIs have become highly sensitive and flexible tools for the functional analysis of genetically defined neural circuits across diverse spatial and temporal scales (Mank et al., 2008; Wallace et al., 2008; Tian et al., 2009; Lütcke et al., 2010). In future, we expect to see novel genetically-encoded probes of neural activity, generated through a combination of rational design and directed mutagenesis.

## 2.3. Delivery of activity indicators to brain

Over the last decade, a diverse set of techniques has been developed to efficiently label neurons and neuronal populations in the living brain with the ever-growing toolbox of functional indicators, comprising both small organic dyes or proteins. Different techniques can be classified according to the number of neurons labelled, the specificity of labelling, the persistence of labelling, and the duration as well as the invasiveness of the procedure. Furthermore, small-molecule indicators and genetically encoded probes in general require different modes of delivery into the CNS.

*Delivery of small-molecule indicators*. Different from lipophilic voltage-sensitive dyes that are incorporated in cell membranes and therefore can be topically applied to the brain surface for staining, most organic calcium indicator dyes are water-soluble and thus require a different loading technique. In the first demonstration of in vivo two-photon imaging of dendritic calcium dynamics in the mammalian brain (Svoboda et al., 1997), as well as in a number of subsequent studies (reviewed in Helmchen and Waters, 2002), single-cell labelling with calcium indicator was achieved by direct injection of the dye into the soma using intracellular recording electrodes followed by passive dye diffusion throughout the dendritic tree. This challenging loading technique allows for activity measurements in dendritic branches and spines, which also can be related to the simultaneous recording of spiking activity or sub-threshold membrane potential dynamics (Waters et al., 2003; Waters and Helmchen, 2004; Jia et al., 2010). While intracellular loading is beneficial for studying subcellular calcium dynamics, the technique is less suited for simultaneous labelling of several neurons and thus precludes network analysis. Other techniques that at least in part enable labelling of neuronal populations include fibre-tract loading with dextran-conjugated calcium-dyes followed by antero- or retrograde transport (O'Donovan et al., 1993; Wachowiak and Cohen, 2001) as well as dye electroporation carried out either targeted to particular cells or in a diffuse manner, staining subsets of neurons (Nagayama et al., 2007; Nevian and Helmchen, 2007; Kitamura et al., 2008).

The most common method for labelling hundreds of neurons and astrocytes simultaneously involves pressure injection of AM-ester dyes directly into the brain region of interest (figure 1, Stosiek et al., 2003). AM dyes are membrane-permeable and diffuse passively into neurons and astrocytes. Within cells, endogenous esterases cleave the AM groups so that indicators become fluorescent and membrane-impermeable and are trapped within cells. This staining approach, termed 'multi-cell bolus loading' (MCBL, Stosiek et al., 2003), has been successfully applied with a number of calcium indicators, including Fura-2 AM, Fluo-4 AM, Indo-1 AM and OGB-1 AM, typically staining hundreds of cells within a diameter of several hundred micrometers around the injection site (Stosiek et al., 2003; Kerr et al., 2005). Furthermore, since the dye injection procedure takes only a few minutes, multiple injections can be performed to increase the number of labelled neurons even further. Over the last 8 years, MCBL has been adopted by numerous laboratories and has become the most common technique for in vivo staining of neuronal networks. MCBL has been for example applied in different cortical regions in different species, such as mice, rats, ferrets and cats (Ohki et al., 2005; Ohki et al., 2006; Kerr et al., 2007; Sato et al., 2007; Li et al., 2008; Kara and Boyd, 2009; Stettler and Axel, 2009; Rothschild et al., 2010), and in rodent cerebellum (Sullivan et al., 2005) and spinal cord (Johannssen and Helmchen, 2010). Although MCBL is experimentally straightforward, the approach still suffers from a number of shortcomings as mentioned above. Most importantly, relabeling of the same populations of cells with organic dyes such as OGB-1 has been achieved only recently (Andermann et al., 2010) and seems to work only over a relatively short timeframe (a few days). In most cases an 'acute' surgery is performed to gain access to the brain, followed by dye loading and - within minutes to a few hours - imaging of neural activity. Imaging of the same populations over the course of weeks and months to study long-term changes in neuronal activity during development, learning, plasticity or disease progression, has not been practicable using MCBL.

*Delivery of protein-based activity sensors.* Small-molecule indicators of neural activity are usually applied to the tissue as the final (or near-final) fluorescent product such that dye loading and imaging are performed within close temporal succession of each other. This procedure differs fundamentally for genetically encoded indicators for which the final fluorescent sensor (a protein) is not directly applied to the tissue of interest. Instead, cells are provided with the genetic information (usually in the form of DNA) that codes for the sensor and the cell-intrinsic transcription and translation machinery is 'hijacked' to produce the final protein. Adequate expression levels of proteins are usually only achieved after at least a few days, so that DNA delivery and imaging are performed in different experimental sessions.

Similar to small-molecule indicators, DNA plasmids can be directly electroporated into cells by brief application of electric pulses. To achieve efficient transfection of cells, plasmid electroporation is commonly performed in utero (Saito and Nakatsuji, 2001). In this procedure, small volumes of DNA plasmids are injected into the ventricle of the embryo. Upon directed application of electric pulses, negatively charged DNA plasmids are taken up by neuronal progenitors lining the ventricular walls. When these cells migrate into the cortex, they and their descendant neurons express the transfected DNA. In utero electroporation has been applied to express the genetically encoded calcium indicator TN-XXL in the neocortex of mice (Mank et al., 2008). Because the cortex is built in a regular inside-out manner during development, in utero electroporation at different gestational stages specifically labels neuron in certain cortical layers. For example, following electroporation at embryonic day 16, construct expression is mainly observed in cortical layers 2/3 (Petreanu et al., 2007).

While in utero electroporation provides a convenient and simple way for expression of indicator proteins in the neocortex, this method is limited for in vivo neural network imaging because of the low fraction of neurons expressing the protein (max. 20%, depending on cortical area; Petreanu et al., 2007; Adesnik and Scanziani, 2010). An alternative method that achieves much higher transfection rates is viral vector mediated delivery of genes. Viruses are essentially encapsulated genetic material that can infect cells and use cells' transcription and translation machinery to express viral genes. In the context of gene therapy, several recombinant vectors have been engineered that allow for stable gene expression over long time periods without toxicity (Verma and Weitzman, 2005). The most commonly used recombinant viral vectors for gene delivery into the brain are the DNA virus herpes simplex virus (HSV) and adeno-associated virus (AAV) as well as RNA-based lentivirus. Numerous pseudotypes of recombinant viral vectors can be created by exchange of native structural virus proteins with those from some other virus, which allows modification of the viral tropism for certain cell types. Recently, viral transfection by AAV has emerged as the method of choice for the delivery of various GECIs into the mouse neocortex (Wallace et al., 2008; Tian et al., 2009; Dombeck et al., 2010; Lütcke et al., 2010). AAV-GECIs can be stereotactically injected into the brain region of interest in a simple and quick surgery. Two to three weeks after injection, high levels of expression are observed in a large fraction of cells (up to 100%; Dombeck et al., 2010) up to 1 mm from the injection site (Lütcke et al., 2010) (figure 2). Importantly, expression of GECIs remains stable for weeks to months thus allowing for long-term, chronic recordings of neural network activity (Tian et al., 2009). In addition to their tropism for certain cell types, viruses allow for circuit level specificity of transgene expression because they can efficiently infect neurons through their axon terminals. Using this approach, subpopulations of neurons that project their axons into a certain target region can be selectively labelled with indicators.

While virus-mediated delivery has become a highly versatile and simple expression strategy that is, in principle, not limited to a given species, the generation of transgenic animals remains in many ways the 'holy grail' for the expression of genetically encoded activity reporters. Transgenic expression of indicators would be particularly beneficial for studying the development of neural activity even a few days after birth, something that is not possible using viruses due to the prolonged expression period. So far, the production of transgenic mice expressing a number of different GECIs has been reported (Hasan et al., 2004; Heim et al., 2007) including membrane-bound YC3.60 (Nagai et al., 2004). In one of these studies, sensory-evoked $Ca^{2+}$ transients were demonstrated using wide-field imaging of the olfactory bulb with two different GECIs expressed under a tetracycline-inducible promoter (Hasan et al., 2004). The widespread use of these mouse lines has been limited, however, by the failure to clearly monitor AP-evoked cellular $Ca^{2+}$ signals in vivo, possibly due to low protein expression levels. It remains to be seen if transgenic mouse lines with boosted expression of the newest generation calcium and voltage sensors will match virus-based expression systems in terms of sensitivity to neural activity and stability of expression.

*Achieving cell type specificity.* It has been appreciated for a long time that neocortical circuits are composed of different neuronal cell types with the hypothesis that different types of neurons are performing different functions within the circuit. At the most basic level, excitatory pyramidal cells can be distinguished from inhibitory GABAergic interneurons. Moreover, it is becoming increasingly clear that a considerable diversity in terms of biochemistry, physiology and connectivity exists within interneurons (Ascoli et al., 2008). For example, basket and chandelier interneurons are known to form synapses on the perisomatic region of pyramidal cells, thus closely controlling their output, whereas double bouquet cells synapse on distal dendritic arborizations and therefore modulate synaptic inputs (for review see Douglas and Martin, 2009). Importantly, these interneuron subtypes can be distinguished also by the expression of different calcium binding proteins, with basket and chandelier cells expressing parvalbumin whereas double bouquet cells express calbindin and calretinin. Accounting for the diversity of cell types in neocortical circuits is a challenge that has only recently started to be addressed by in vivo functional imaging studies.

Perhaps the most straightforward means of achieving some kind of specificity is the use of cell type specific promoters. For example, GABAergic interneurons can be visualized by expressing GFP under control of the glutamic acid decarboxylase (GAD) promoter in transgenic mouse lines (Tamamaki et al., 2003). Since GFP can be spectrally separated from calcium indicators, this strategy has been used to perform two-photon calcium imaging in defined interneurons and relate their response profiles to those of pyramidal cells (Sohya et al., 2007; Gandhi et al., 2008). In a similar vein, GECIs may be expressed under specific promoters, such as the neuron-specific synapsin promoter, to target their expression specifically to neurons (Tian et al., 2009; Lütcke et al., 2010). Because GECI expression requires very strong promoters, however, this strategy does not seem promising as means of more fine grained distinction between subtypes of neurons. Instead, conditional gene expression strategies, developed by molecular biologists, will be potentially powerful tools for targeting expression of genetically encoded sensors of neural activity to certain cell types. Since a comprehensive discussion of conditional gene expression systems is beyond the scope of this review (see Luo et al., 2008 for review) we focus on the Cre-loxP system which arguably will be of most immediate benefit for neural circuit analysis. The Cre-loxP system (reviewed in Nagy, 2000) achieves a high level of cell type specificity by combination of two different genetic manipulations. First, a transgene of interest is delivered to the brain, for example by viral injection, as described above. Importantly, expression of the transgene is prevented by insertion of a transcription stop signal flanked with loxP sites before the DNA sequence coding for the reporter. In a second step, the DNA recombinase Cre is expressed under control of a cell-type specific promoter. Cre recognizes loxP sites and causes a site-specific recombination event that removes the stop signal leading to transcription of the transgene only in Cre-expressing cells. Recently, this approach has been used to express the GECI GCaMP3 selectively in parvalbumin interneurons by injection of Cre-dependent AAV-GCaMP3 in PV-Cre mice (Tian et al., 2009). The promise of this approach for neuroscience stems from the large number of different Cre driver lines that are being produced in projects such as the Gene Expression Nervous System Atlas (http://www.gensat.org; Gong et al., 2007) or the Allen Institute for Brain Science Atlas (http://mouse.brain-map.org). In the future, even more precise control over GECI expression may be achieved by temporal regulation of transcription using the tamoxifen-inducible CreER system (Feil et al., 1996).

Instead of selectively labelling and imaging only a certain population of neurons, an alternative approach may be to stain all neurons (using synthetic or genetically encoded indicators) and determine the exact cell type post-hoc using immunohistochemistry procedures. This approach is challenging because images acquired in vivo have to be co-registered with immunohistochemically processed sections wherein corresponding cells must be identified. Recently, Kerlin et al. (2010) performed two-photon calcium imaging in the visual cortex of transgenic mice expressing GFP in interneurons; after the imaging sessions, the brains were sliced and immunostained for 3 different classes of interneurons (Kerlin et al., 2010). After successful co-registration of immunostained slices and in vivo imaging data, it was possible to distinguish functional responses of four different neuron subtypes (parvalbumin-expressing interneurons, somatostatin-expressing interneurons, vasoactive intestinal peptide-expressing interneurons, and excitatory, GAD-negative neurons). In contrast to pyramidal cells, none of the interneuron subtypes showed any orientation selectivity in response to stimulation by drifting gratings (Kerlin et al., 2010). In summary, we anticipate that some if not all of these approaches for cell type discrimination will be applied much more extensively in future studies and should soon yield important new insights into the functional organization of neural circuits.

## 3. Analysis of neural network dynamics

In the previous sections, we have highlighted how recent technological advances made it possible to look at the function of neuronal populations in ways that were largely unthinkable only a few years ago. Moreover, upcoming improvements in microscopy techniques and indicator design are likely to expand these opportunities even further. While these imaging techniques provide neuroscientists with unprecedented insights into neural network dynamics, they also pose increasing challenges for the analysis of the large data sets acquired. In the following section we will discuss analytical strategies that have been developed to meet these challenges, especially focussing on the analysis of two-photon calcium imaging data. We will conclude with a discussion on how the new types of experimental data might be combined with modelling approaches to further our understanding of neural circuits.

### 3.1. Preprocessing and segmentation of calcium imaging data

In a typical calcium imaging experiment, fluorescence intensities are sampled repeatedly from a population of neurons stained with a calcium indicator while the animal receives a sensory stimulation or performs a task. Since fluorescence intensity is related to neuronal spiking, an initial analysis goal is the extraction of fluorescence intensity over time for each neuron. To segment neuronal cell bodies from unspecific background staining, most previous studies have relied on time-consuming manual segmentation based on anatomical criteria. This approach scales very poorly with the expected increase in the size of populations imaged, therefore necessitating the development of semi-automatic or fully automatic algorithms for segmentation of calcium imaging data. We have recently developed a semi-automatic strategy for the segmentation of neuronal somata labelled with the GECI YC3.60 (Lütcke et al., 2010) that requires users to mark only the approximate position of cell bodies in the image frame. The shape of somata in the vicinity of marked seed points is then determined by an active contour segmentation routine (Li and Acton, 2007). Alternatively, regions of interest (ROIs) can be segmented from the image according to functional criteria (Ozden et al., 2008; Junek et al., 2009; Mukamel et al., 2009; Valmianski et al., 2010; Miri et al., 2011). Recently, this approach has been fully automated for two-photon calcium imaging data acquired in the cerebellum of awake mice (Mukamel et al., 2009). After dimensionality reduction and noise removal by principal component analysis (PCA), correlated cellular signals were identified by spatio-temporal independent component analysis (ICA). In a last step, neurons and astrocytes with correlated calcium traces were separated by morphological considerations. While automated image segmentation based on correlated functional signals provides a fast and efficient strategy for analysis of large-scale calcium imaging data sets, it is currently limited to acquisitions with very good signal-to-noise ratio (see also Dombeck et al., 2010).

Moreover, the approach only identifies 'active' cells thus making it impossible to quantify the proportion of non-responsive neurons in a population.

Due to the high spatial resolution of two-photon microscopy, even small movements (a few micrometers) can cause artefacts during image acquisition. Moreover, because of the optical sectioning properties of two-photon microscopy, neurons moving in and out of the focal plane can even cause apparent calcium transients that may be mistaken for neuronal activity. Fortunately, mechanical tissue stabilization using agar and transparent rubber can largely eliminate these problems, at least in anaesthetized preparations. Moreover, the use of ratiometric indicators reduces the risk of spurious transients as motion affects both channels equally and therefore cancels out when computing the ratio. Imaging in awake, behaving animals frequently results in residual motion, which can be corrected for using offline algorithms based on Hidden-Markov models (Dombeck et al., 2007) or Lucas-Kanade image registration (Greenberg and Kerr, 2009).

Population calcium imaging usually focuses on somatic calcium signals which can be employed to estimate AP firing. Nevertheless, most labelling techniques also stain neuronal processes such as dendrites, which are usually too thin to be resolved by calcium imaging. Non-somatic labelling gives rise to diffuse background staining which shows calcium-dependent changes in fluorescence due to concerted activation of neuronal populations for example in response to sensory stimulation. While the so-called neuropil signal has been used as kind of an optical equivalent of the local field potential (Kerr et al., 2005), it may also contaminate measurements from somatic ROIs. This problem is particularly severe when stimulation drives a large neuropil signal and average stimulus-locked calcium transients in cells are analyzed. The degree of neuropil contamination depends on the effective numerical aperture (NA) of the objective, decreasing at higher NA (Göbel and Helmchen, 2007a; Kerr and Denk, 2008). To correct for neuropil contamination it is also possible to estimate the fluorescence surrounding a cellular ROI and subsequently adjust the relative change in fluorescence by the expected neuropil contamination based on the effective NA.

3.2. Reconstruction of neuronal firing from calcium signals

Following these pre-processing procedures, spiking activity in neurons must then be reconstructed as accurately as possible. Calcium imaging reports cellular calcium concentration changes, which often can be directly linked to supra-threshold neural activity as described above. During spike trains, successive calcium transients evoked by the APs sum linearly for calcium concentrations far from indicator saturation. Consequently, the fluorescence trace can be considered a simple convolution of the underlying spike train with a template single AP-evoked calcium transient function, e.g., according to equation 3 or 4 (Helmchen, 2011). In principle, then, fluorescence traces can be deconvolved by inverse filtering with a template kernel to yield the underlying firing pattern of the recorded neurons (Wiener filter; Yaksi and Friedrich, 2006; Holekamp et al., 2008). Unfortunately, linear deconvolution is highly sensitive to noise in the calcium traces (i.e. deviations from the filter kernel). To account for the high noise sensitivity of the Wiener filter, Vogelstein et al. (2010) have recently proposed a deconvolution algorithm that restricts inferred spike trains to be positive and thus is able to determine the most likely firing pattern underlying observed fluorescence data. An advantage of the algorithm is its efficient implementation, which allows real-time spike train reconstruction from large neuronal populations (> 100 neurons). While deconvolution-based approaches are able to determine changes in AP firing pattern underlying calcium indicator fluorescence traces, it is frequently also desirable to estimate precise spike timings for individual APs. A number of different approaches have been proposed to achieve this goal, especially within sparse firing regimes commonly observed in neocortical areas. Initially, simple thresholding and template-matching strategies have been employed to find neuronal firing events. While these strategies are simple and intuitive, they lack the capability to determine the precise number of APs per event and also suffer from poor sensitivity, compared to more advanced algorithms. On the other hand, Sasaki et al. (2008) employed a supervised classification approach to determine spike times from fluorescence data with high reliability. In practice, machine learning approaches are limited, however, because large example data sets with joint electrophysiological and optical recordings are required to train the algorithm. Recently, we have combined sophisticated thresholding and template matching algorithms and devised an iterative 'peeling' strategy for reconstruction of spike times from high-speed calcium imaging data (Grewe et al., 2010). In this approach, epochs of neuronal firing were detected using a modified Schmitt trigger thresholding procedure, which searches through the fluorescence trace for periods that first pass a high threshold and then stay elevated above a second threshold for at least a minimum duration (Grewe et al., 2010). For each event the occurrence of an AP is presumed and the stereotyped single AP-waveform is iteratively subtracted from the fluorescence trace until no more events are detected and a noisy residual trace remains. In this study, we presumed a single AP-waveform according to equation 4 and parameters were derived by fitting calcium transients, which were verified to stem from single APs by electrophysiological recordings under identical experimental conditions. In combination with high-speed imaging, this approach allowed high-fidelity reconstruction of spike times from calcium traces with near-millisecond temporal precision (Grewe et al., 2010).

The performance of reconstruction algorithms depends on the acquisition rate and the signal-to-noise, defined as the single AP calcium transient amplitude divided by the standard deviation of the noisy baseline fluorescence. This is exemplified with simulated fluorescence traces in figure 3. The analysis of detection accuracy (and false-positive rate) indicates that a high temporal resolution of >30 Hz and a SNR of at least 3 are beneficial for a good estimate of the spike train. A major limitation of template-based reconstruction algorithms is, however, their assumption that all neurons in the recorded population conform to the same single AP-waveform. While this assumption is probably valid for neocortical pyramidal neurons, other cell types - in particular some sub-types of inhibitory interneurons - will certainly exhibit different calcium dynamics (Helmchen and Tank, 2011). In the future, this will need to be taken into account. Because many fast-spiking neocortical interneurons elicit APs at higher rates compared to pyramidal neurons, determination of precise spiking pattern may not be possible for these cells. Under these circumstances, it may however still be feasible to estimate neuronal firing rates using deconvolution-based approaches (Yaksi and Friedrich, 2006; Moreaux and Laurent, 2008).

## 3.3. Analysis of network activity

Following signal normalization and preprocessing, as well as reconstruction of neuronal firing events, two-photon calcium imaging data can be described as population raster plots, similar to classical electrophysiological recordings, but typically with higher dimensionality because local neuronal populations comprising tens to hundreds of neurons are sampled. Several calcium imaging studies have measured functional properties of individual neurons, such as their tuning in response to a stimulus, and related these to the location of cells within the local population. In the simplest case, average responses for two stimuli (e.g. visual gratings of different orientation or tones of different frequencies) are determined for each cell and the relative difference in response strength is computed as a measure of response selectivity (selectivity index, SI):

$$SI = \frac{R_1 - R_2}{R_1 + R_2} \tag{5}$$

where $R_1$ and $R_2$ refer to the average response strength for stimulus 1 and 2, respectively. Spatial maps of cellular tuning properties, obtained by color-coding SI, have demonstrated that functional properties in the visual cortex of higher mammals are arranged with remarkable precision (Ohki et al., 2005; Ohki et al., 2006; schematized in figure 4; Kara and Boyd, 2009). On the other hand, many cortical areas, especially in rodents, lack a clear topographic organization on the level of local neuronal populations (Ohki et al., 2005; Kerr et al., 2007; Sato et al., 2007; Stettler and Axel, 2009; Bandyopadhyay et al., 2010; Rothschild et al., 2010). More subtle spatial dependencies of functional properties between neurons may be revealed by clustering techniques such as k-means clustering (Dombeck et al., 2009) or nearest-neighbor analysis.

While static neuronal response properties have been extensively investigated using in vivo two-photon calcium imaging, it remains a challenge for the field to characterize the dynamic interactions among neurons in the microcircuit and thereby obtain a more complete description of emergent computational properties of local neuronal populations. In the following section we briefly describe analytical techniques that have recently been employed to describe dynamical network properties based on calcium imaging data. We will close by a brief discussion of how these data-driven approaches may in the near future inform theoretical concepts of cortical network function.

At the most basic level, interactions in local populations may be quantified using pair-wise correlation of response patterns between individual neurons. The correlation coefficient r between two variables (i.e calcium traces from two neurons) is defined as the covariance of the two vectors (X, Y), normalized by their respective variability (standard deviation, $\sigma_X$ and $\sigma_Y$):

$$r_{xy} = \frac{\text{cov}(X,Y)}{\sigma_x \cdot \sigma_y} \qquad (6)$$

The overall degree of interaction within a network is described by a matrix of pairwise correlation coefficients between all neurons. Correlations between two neurons may refer to the similarity of their average stimulus-evoked responses, as defined for example by the peristimulus time histogram (PSTH). Using calcium imaging, these 'signal' correlations were found to be relatively weak ($< 0.2$) and to decrease with distance between neurons in different cortical areas, including barrel cortex (Kerr et al., 2007) and auditory cortex (Rothschild et al., 2010).

While 'signal' correlations describe the overall similarity of neuronal responses, 'noise' correlations quantify the trial-to-trial co-activation of two neurons. High noise correlations may indicate that two neurons are directly connected or share common inputs, although the low temporal resolution as well as limited trial number in calcium imaging experiments usually precludes direct conclusions about the underlying connectivity. On average, noise correlations between nearby neurons tend to be higher than signal correlations and to decrease with distance in auditory cortex (Rothschild et al., 2010) and barrel cortex (Kerr et al., 2007). Similarly noise correlations that decrease with distance between neurons are also observed during spontaneous activity (Kerr et al., 2007; Rothschild et al., 2010). Notably, the low correlations determined by imaging studies are supported by numerous electrophysiological investigations which have found low (Zohary et al., 1994; Lee et al., 1998; Constantinidis and Goldman-Rakic, 2002; Gutnisky and Dragoi, 2008) or even near-zero (Ecker et al., 2010) correlations in the mammalian neocortex. This contrasts with strong neural synchrony observed in other systems such as the mammalian hippocampus (Harris et al., 2003), songbird high vocal center (Hahnloser et al., 2002) or spinal cord pattern generators (Kwan et al., 2009).

Recently, two-photon calcium imaging in awake mice performing a task provided evidence that these correlations are probably behaviourally relevant (Komiyama et al., 2010). In this study, stronger correlations between neurons involved in the choice behaviour were observed as mice learned the task and improved in performance. Interestingly, representational sparseness also increased with task performance, suggesting that the sparse coding scheme in neocortex reflects a high degree of functional specialization. Although pairwise correlations can be determined easily, the non-random connectivity in local populations (Song et al., 2005) suggests that higher order interactions (between three or more neurons) may contribute significantly to the functional organization of local neuronal populations. In fact, recent multi-electrode recordings in the visual cortex of anaesthetized monkeys highlight the importance of higher-order interactions especially in local clusters of cells (Ohiorhenuan et al., 2010). Unfortunately, analyzing higher-order correlations within typical neuronal populations is challenging due to the combinatorial explosion of possible interactions (which grows with the factorial of population size). Leaving computational considerations aside, it is impractical to collect sufficient data in order to assess all connections even in moderate sized populations (e.g. an exhaustive analysis of a 10-neuron network would require a minimum of $10! = 3628800$ combinations to be tested, even leaving aside issues such as state-dependent modulations of synaptic connections).

A common approach for reducing the complexity of high-dimensional data sets is the use of dimensionality reduction techniques, such as principal component analysis (PCA) or multidimensional scaling. PCA is particularly suited for the analysis of multi-cell imaging data, because responses from individual cells are frequently correlated. Principal component analysis reduces the dimensionality of the original data set by orthogonal transformation into uncorrelated variables (components). If the cumulative explained variance for the first 2-3 principal components is high, these can be considered as an adequate, lower-dimensional description of the high-dimensional network state (figure 4). Recently, Niessing and Friedrich (2010) employed PCA to analyze the temporal evolution of odor-evoked activity patterns measured by two-photon calcium imaging in the olfactory bulb of zebrafish (Niessing and Friedrich, 2010). By projection of population activity patterns onto the first three principal components, it was possible to analyze the trajectories of activity patterns in principal component space after stimulation with different odors (Niessing and Friedrich, 2010). While trajectories diverged over time for different odors, stimulation with the same odor at different concentrations resulted in converging trajectories in principal component space. Thus PCA allows identification of unique network states, potentially coding for different stimuli.

While dimensionality reduction techniques such as PCA extract relevant information from population activity data, decoding strategies attempt to classify the stimulus-specific information contained in the activity patterns. In the case of two stimuli or conditions, a classification algorithm finds a hyperplane that splits the high-dimensional data space and classifies values on different sides of the hyperplane into the two categories. To avoid over-fitting, the classification rule is applied to an independent data set and above-chance classification is taken as evidence that stimulus-specific information is contained in the input data. This approach naturally extends to the high-dimensional data acquired with two-photon calcium imaging. Decoding approaches can provide insights into the degree of stimulus-specific information carried by neuronal ensembles of different sizes or at different locations within the brain. In the rat barrel cortex, for example, the occurrence of whisker deflections could be decoded with higher accuracy from the activity patterns of larger populations, compared to smaller networks or single cells (Kerr et al., 2007). Similar analyses can be performed in the temporal domain, in order to determine when and for how long stimulus-specific information is contained in local populations. Although dimensionality-reduction and classification analysis strategies can be performed independently of each other, it is frequently advantageous to combine both approaches. Decoding analysis can be problematic, for example, if the number of dimensions is much larger than the number of measurements, which is becoming realistic for population imaging experiments (where hundreds of cells may be recorded simultaneously). To circumvent this problem, important features can be extracted from raw data using reduction techniques, for example PCA. Subsequently, classification or discrimination analysis techniques may be applied to the reduced feature space.

We expect that in the coming years, high-speed imaging approaches in combination with advanced data analysis techniques will eventually permit a description of the temporal propagation of stimulus-evoked activity through the cortical network with various cell types discriminated. These experiments will allow us to test and extend current models of cortical information processing. State-dependent network models hypothesize that information is encoded in the evolution of trajectories in high-dimensional space formed by the activity vectors of many neurons. On the other hand, attractor models describe cortical computations as steady-state attractor points with constant neuronal firing rates. Using experimental approaches described in this review, it should soon become possible to directly observe the dynamics of neuronal activity patterns in response to sensory stimuli or preceding motor behavior and thereby evaluate different theoretical models of cortical function.

## 4. Outlook

Over the past 50 years, extracellular recording studies have provided valuable information on the functional properties of neurons in many cortical areas, both in anaesthetized and awake, behaving animals. In these studies, investigators blindly advance the recording electrode through the cortical tissue, while at the same time 'listening' for neurons with high firing rates. Using this technique, it is therefore not possible to obtain an unbiased sample of the distribution of responses in local neuronal populations. Imaging approaches, such as two-photon calcium imaging, sample from most or all cells within a population in a given field-of-view. It therefore becomes possible to obtain an unbiased estimate of the distribution of functional properties within a local population of neurons. It has indeed emerged over the last years, largely based on evidence from imaging studies, that neuronal properties in many cortical areas are highly heterogeneous and non-uniformly distributed. Studies in the rodent barrel cortex as well as in auditory cortex have revealed that the majority of neurons respond to sensory stimulation with very low rates of action potential firing, with a substantial fraction of cells even being 'silent' (i.e. never responding under the tested conditions). Neuronal populations in cortical layer 2/3 therefore frequently operate in a regime of sparse coding, both in anaesthetized and awake animals. On the other hand, small subsets of highly active neurons apparently exist, which may have special functions in signal transmission. Imaging studies have provided evidence for a highly skewed distribution of firing rates in neocortical layer 2/3. These findings have been corroborated by unbiased electrophysiological recordings in anaesthetized and awake animals and are in line with the non-random functional and anatomical connectivity between neurons in local populations. Further improvements in microscope techniques should allow imaging of even larger networks of cells, not only in-plane but also in 3D. This will allow more exhaustive sampling of neurons in local networks, thus permitting identification and characterization of rare high-responsive cells. On the other hand, high-speed imaging using for example AOD-based approaches will allow for more accurate measurements of correlations within neocortical networks, even on a millisecond timescale. In combination with long-term expression of GECIs, it will become possible to investigate the stability of functional cellular and network properties over extended time periods as well as their alteration during learning, plasticity or even in diseases. While the majority of imaging studies thus far has been performed in anaesthetized animals (Ohki et al., 2005; Ohki et al., 2006; Kerr et al., 2007; Sato et al., 2007; Kara and Boyd, 2009; Stettler and Axel, 2009; Rothschild et al., 2010), anaesthesia precludes the investigation of neural circuits during motor or task-driven behaviours, such as decision making. Recently, a number of groups have achieved two-photon calcium imaging in awake mice and rats using both synthetic and genetically encoded calcium indicators (Dombeck et al., 2007; Dombeck et al., 2009; Nimmerjahn et al., 2009; Sawinski et al., 2009; Andermann et al., 2010; Dombeck et al., 2010; Komiyama et al., 2010). These approaches will allow identification of the specific neural circuitry underlying complex behaviours and, in combination with the emerging optogenetic toolbox (Deisseroth, 2011), manipulation of these circuits to firmly establish a causal relationship between neural activity and behaviour.

A major limitation of current imaging approaches is the lack of information regarding the underlying connectivity between the imaged cells (the 'wiring diagram'). Reliable identification of synaptic contacts and fine neuronal processes requires a spatial resolution in the range of tens of nanometers, which is beyond the reach of conventional optical approaches. Besides novel super-resolution optical methods (Hell, 2007), electron microscopy (EM) remains the only technique capable of complete anatomical reconstruction of neural circuits (Briggman and Denk, 2006). In recent years, the time consuming process of serial-section EM has been sped up by the introduction of block-face EM techniques (Denk and Horstmann, 2004) which allow automated acquisition of small volumes of neural tissue (Briggman and Denk, 2006). These advances have recently allowed for functional two-photon calcium imaging in the retina and visual cortex, followed by post hoc EM reconstruction of the underlying anatomical circuitry between imaged neurons (Bock et al., 2011; Briggman et al., 2011).

Given further improvements in the speed of EM acquisition as well as development of fully automated algorithms for tracing of neural structures (Jurrus et al., 2009; Turaga et al., 2010), future imaging experiments may routinely be combined with post hoc volume-based EM reconstruction to reveal the full underlying wiring diagram and thereby link structural and functional information. Finally, all these experimental innovations will require the concomitant development of novel analytical tools and theoretical approaches to neuronal networks, which will eventually lead to true advances in the understanding of cortical circuits.

## 5. Conclusions

Neural circuits are complex dynamic networks par excellence, which for many decades have appeared almost impenetrable. In this review we have described novel methods that have emerged only recently and that we have now at hand to collect comprehensive and quantitative data from local neural networks in living animals. The toolbox for accurate network measurements is still growing, with microscopy techniques pushing beyond their current limits, with new indicators of activity being designed, and with genetic means being exploited for targeting of specific subsets of cells. Expecting large data sets of in vivo neural network dynamics in the coming years, it will be essential to optimize and automatize analysis techniques, taking advantage of newest computer technology. Experimental data sets on network dynamics during behaviour should be particularly valuable as they will make it possible to support or falsify current theoretical models of neural circuit function. Given the tremendous speed of research progress it now seems a realistic task to obtain a mechanistic understanding of signal flow through neuronal microcircuits and its operational principles.

**Acknowledgments**

We thank Jerry Chen for comments on the manuscript. The authors acknowledge support from the German Academic Exchange Service (DAAD) to H.L., and from the Swiss Systems Biology Initiative SystemsX.ch (project Neurochoice), and the EU-FP7 programme (project 243914, BRAIN-I-NETS) to F.H.

**Figures**

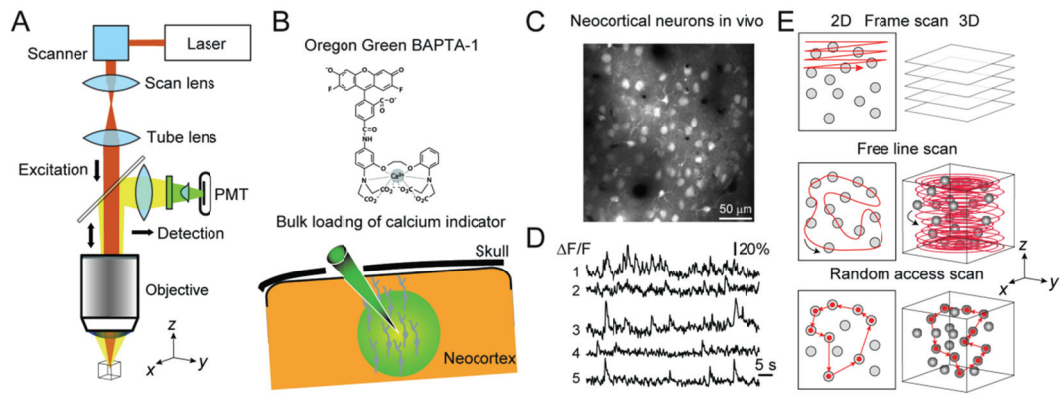

**Figure 1**. Two-photon imaging of neuronal populations in vivo. (A) Generic scheme of a two-photon laser scanning microscope. The laser beam is focused through a microscope objective and scanned to sample fluorescence from either an *xy*-plane or a volume in the tissue. (B) Loading of neocortical cells with the synthetic calcium indicator dye Oregon Green BAPTA-1 by direct AM-dye injection. (C) Example in vivo two-photon fluorescence image of a cell population in layer 2/3 of mouse neocortex obtained with a standard *xy*-scan. Cells filled with Oregon Green BAPTA-1 comprise neurons as well as astrocytes (typically brighter). (D) Example fluorescence traces from 5 neurons in a different population, showing action potential-evoked calcium transients. Fluorescence changes are expressed as percentage changes relative to baseline (ΔF/F). (E) Various 2D and 3D laser scanning approaches can be used to optimize speed and/or number of cells for measurements of neuronal network dynamics. Standard frame scanning approaches are rather slow. Free line scans and random access scanning enable higher acquisition rates. Application of 3D random access scanning for in vivo calcium imaging is still pending.

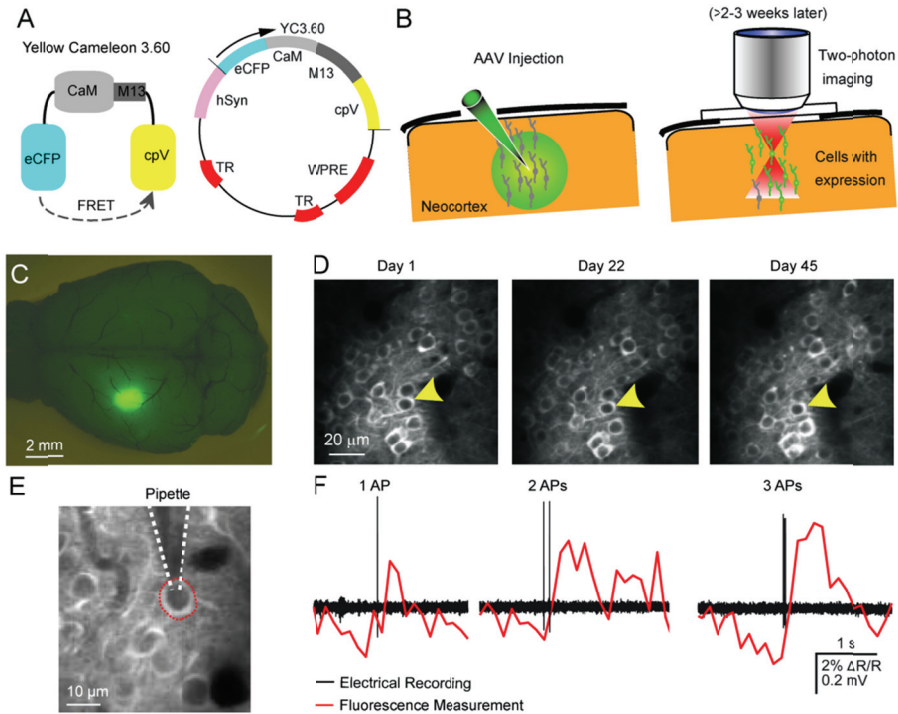

**Figure 2**. Long-term expression and action-potential detection with genetically-encoded calcium indicators (GECI) in the mouse brain. (A) Left: Basic structure of the GECI Yellow Cameleon 3.60 (YC3.60). The indicator consists of two fluorescent proteins (eCFP and cpV), linked by a calcium-sensing domain (CaM). Calcium-binding to CaM brings eCFP and cpV into closer proximity, thus increasing the FRET ratio between the two. Right: YC3.60 is encoded by an adeno-associated viral vector (AAV) under control of the synapsin promoter (hSyn) which allows neuron-specific expression. WPRE … woodchuck posttranscriptional regulatory element, TR … terminal repeats (B) AAV-YC3.60 is slowly injected into the neocortex of anaesthetized mice using a fine glass needle. After 2–3 weeks of expression, cells expressing the calcium indicator fluoresce brightly and can be imaged in vivo using two-photon microscopy. (C) Wide-field fluorescence image of a mouse brain 4 weeks after virus injection. Note the bright fluorescence spot in the left hemisphere indicating the area of YC3.60 expression. (D) Example of neuronal population expressing YC3.60 and imaged for > 1 month using two-photon microscopy. Arrow point indicates a specific neuron identified on all three imaging days. Note that YC3.60 labels mainly somata and dendrites but is excluded from the nucleus. (E) Simultaneous electrophysiological recording with a patch pipette and two-photon imaging for determining the sensitivity of GECIs for AP firing. The image shows a small neuronal population labelled with YC3.60. One neuron has been selected for electrophysiological recording of membrane potential using a recording pipette (highlighted for clarity). (F) Example calcium transients (red) and juxtacellular recordings of neural spikes (black) from the cell shown in (E) for 1, 2 and 3 APs. Even a single spike is accompanied by a clear calcium transient while larger bursts result in calcium transients with increasing amplitude. (C,E,F) modified from (Lütcke et al., 2010).

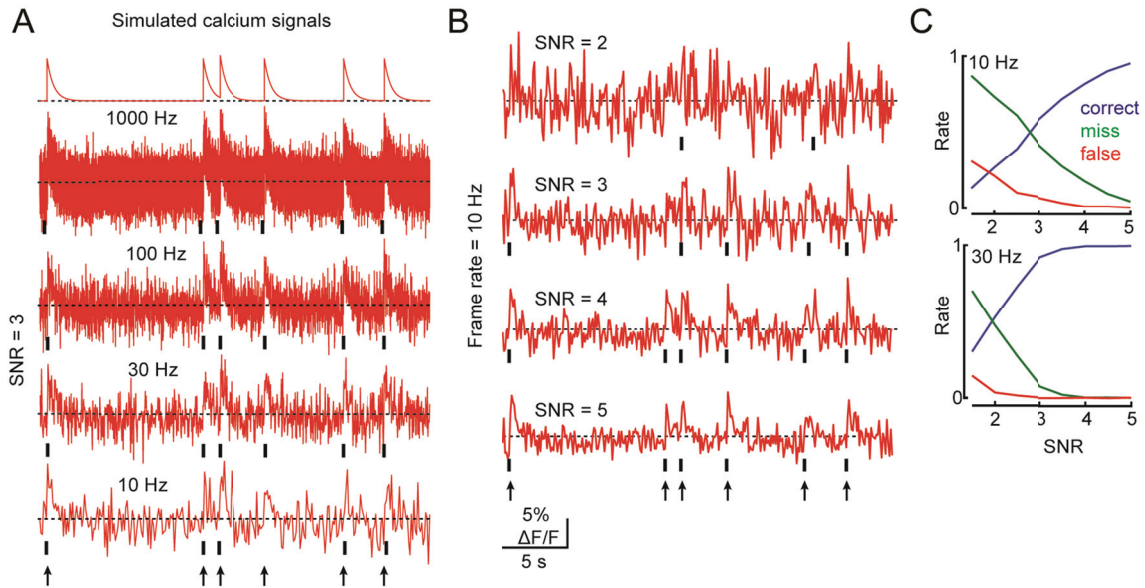

**Figure 3**. Spike reconstruction. (A) Simulated calcium transients in response to action potentials (AP, arrows). Top trace shows the expected calcium dynamics reported by the fluorescence indicator in the absence of noise. Simulation parameters: single-exponential model with A = 7% ΔF/F and τ = 500 ms. Subsequent traces show expected fluorescence in the presence of additional noise (white noise, SNR = 3) sampled at different frame rates. Automatic AP detection (black ticks) using the 'Peeling' algorithm. Even with identical noise levels, calcium transients are more difficult to detect at lower frame rates. (B) Data were simulated as in (A), frame rate is 10 Hz. Whereas APs are faithfully reconstructed by the algorithm at low noise levels, performance deteriorates at higher noise levels. (C) Summary of algorithm performance as a function of SNR at two different frame rates (top: 10 Hz, bottom: 30 Hz). Correct rate refers to the fraction of correct detections (relative to the total number of detected APs) whereas false rate refers to the fraction of incorrect detections (relative to the total number of detected APs). Miss rate refers to the fraction of simulated APs for which no spike was detected within 0.5 s (relative to the total number of simulated APs).

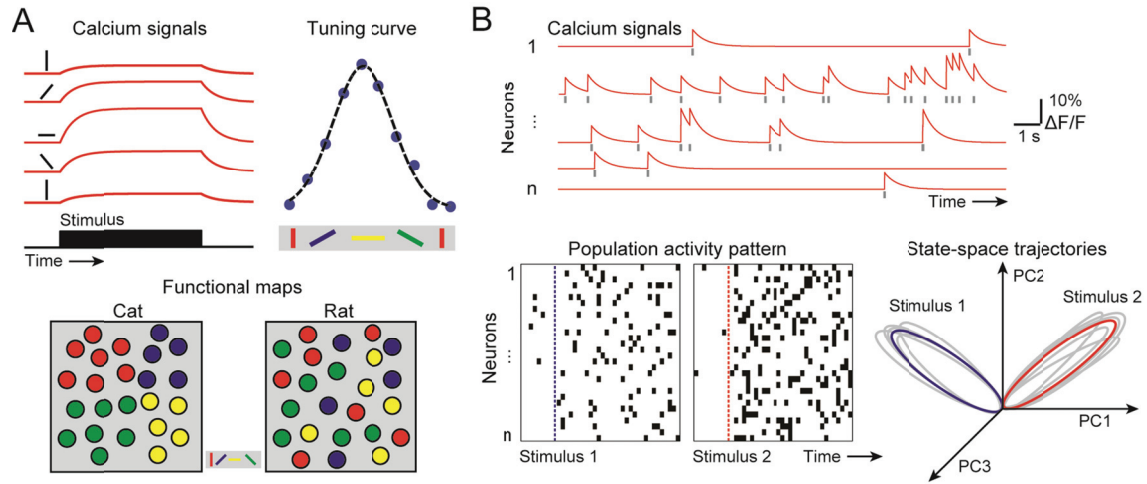

**Figure 4**. Analysis of population calcium signals. (A) Average neuronal tuning properties can be obtained by measurement of calcium signals elicited by different stimuli. In the primary visual cortex, neurons preferentially respond to bars presented at a certain orientation. By fitting the maximum calcium response amplitude for each stimulus, a tuning curve is obtained. Bottom: Colour-coding the preferred orientation for each neuron results in a functional map of orientation-preference at cellular resolution. Whereas neurons with similar orientation preference cluster in the visual cortex of higher mammals such as cats, rodents show a 'salt-and-pepper' organization without obvious clustering (for details see (Ohki et al., 2005; Ohki et al., 2006)) (B) Dynamic analysis of calcium signals. Imaging at high temporal resolution allows the faithful reconstruction of APs (gray) from calcium signals (red). Schematic representation of single-trial population activity pattern in response to two different stimuli. Black squares indicate imaging frames with AP-related calcium transients. In order to reveal the dynamics of population activity in the high-dimensional space of $n$ neurons, principal component analysis is performed on discrete time bins. Subsequently, evolution of the activity pattern evoked by the two different stimuli can be visualized as state-space trajectories in principal component space.

# References

Adesnik H, Scanziani M (2010) Lateral competition for cortical space by layer-specific horizontal circuits. Nature 464:1155-1160.

Akemann W, Mutoh H, Perron A, Rossier J, Knopfel T (2010) Imaging brain electric signals with genetically targeted voltage-sensitive fluorescent proteins. Nat Methods 7:643-649.

Andermann ML, Kerlin AM, Reid RC (2010) Chronic cellular imaging of mouse visual cortex during operant behavior and passive viewing. Front Cell Neurosci 4:3.

Arieli A, Sterkin A, Grinvald A, Aertsen A (1996) Dynamics of ongoing activity: explanation of the large variability in evoked cortical responses. Science 273:1868-1871.

Ascoli GA et al. (2008) Petilla terminology: nomenclature of features of GABAergic interneurons of the cerebral cortex. Nat Rev Neurosci 9:557-568.

Ataka K, Pieribone VA (2002) A genetically targetable fluorescent probe of channel gating with rapid kinetics. Biophys J 82:509-516.

Baker BJ, Mutoh H, Dimitrov D, Akemann W, Perron A, Iwamoto Y, Jin L, Cohen LB, Isacoff EY, Pieribone VA, Hughes T, Knopfel T (2008) Genetically encoded fluorescent sensors of membrane potential. Brain Cell Biol 36:53-67.

Bandyopadhyay S, Shamma SA, Kanold PO (2010) Dichotomy of functional organization in the mouse auditory cortex. Nat Neurosci 13:361-368.

Bock DD, Lee WC, Kerlin AM, Andermann ML, Hood G, Wetzel AW, Yurgenson S, Soucy ER, Kim HS, Reid RC (2011) Network anatomy and in vivo physiology of visual cortical neurons. Nature 471:177-182.

Botcherby EJ, Juskaitis R, Booth MJ, Wilson T (2007) Aberration-free optical refocusing in high numerical aperture microscopy. Opt Lett 32:2007-2009.

Briggman KL, Denk W (2006) Towards neural circuit reconstruction with volume electron microscopy techniques. Curr Opin Neurobiol 16:562-570.

Briggman KL, Helmstaedter M, Denk W (2011) Wiring specificity in the direction-selectivity circuit of the retina. Nature 471:183-188.

Buzsaki G (2004) Large-scale recording of neuronal ensembles. Nat Neurosci 7:446-451.

Carriles R, Schafer DN, Sheetz KE, Field JJ, Cisek R, Barzda V, Sylvester AW, Squier JA (2009) Invited review article: Imaging techniques for harmonic and multiphoton absorption fluorescence microscopy. The Review of scientific instruments 80:081101.

Cheng A, Goncalves JT, Golshani P, Arisaka K, Portera-Cailliau C (2011) Simultaneous two-photon calcium imaging at different depths with spatiotemporal multiplexing. Nat Methods 8:139-142.

Chorev E, Epsztein J, Houweling AR, Lee AK, Brecht M (2009) Electrophysiological recordings from behaving animals--going beyond spikes. Curr Opin Neurobiol 19:513-519.

Constantinidis C, Goldman-Rakic PS (2002) Correlated discharges among putative pyramidal neurons and interneurons in the primate prefrontal cortex. J Neurophysiol 88:3487-3497.

Deisseroth K (2011) Optogenetics. Nat Methods 8:26-29.

Denk W, Svoboda K (1997) Photon upmanship: why multiphoton imaging is more than a gimmick. Neuron 18:351-357.

Denk W, Horstmann H (2004) Serial block-face scanning electron microscopy to reconstruct three-dimensional tissue nanostructure. PLoS Biol 2:e329.

Denk W, Strickler JH, Webb WW (1990) Two-photon laser scanning fluorescence microscopy. Science 248:73-76.

Denk W, Piston DW, Webb WW (2005) Multi-photon molecular excitation in laser-scanning microscopy. In: Handbook of Biological Confocal Microscopy, 2nd edition Edition (Pawley JB, ed). New York: Plenum Press.

Diaspro A, ed (2002) Confocal and Two-Photon Microscopy: Foundations, Applications, and Advances. New York: Wiley-Liss.

Dimitrov D, He Y, Mutoh H, Baker BJ, Cohen L, Akemann W, Knopfel T (2007) Engineering and characterization of an enhanced fluorescent protein voltage sensor. PLoS ONE 2:e440.

Dombeck DA, Graziano MS, Tank DW (2009) Functional clustering of neurons in motor cortex determined by cellular resolution imaging in awake behaving mice. J Neurosci 29:13751-13760.

Dombeck Da, Khabbaz AN, Collman F, Adelman TL, Tank DW (2007) Imaging large-scale neural activity with cellular resolution in awake, mobile mice. Neuron 56:43-57.

Dombeck DA, Harvey CD, Tian L, Looger LL, Tank DW (2010) Functional imaging of hippocampal place cells at cellular resolution during virtual navigation. Nat Neurosci 13:1433-1440.

Donnert G, Eggeling C, Hell SW (2007) Major signal increase in fluorescence microscopy through dark-state relaxation. Nat Methods 4:81-86.

Douglas RJ, Martin KA (2009) Inhibition in cortical circuits. Curr Biol 19:R398-402.

Dreosti E, Odermatt B, Dorostkar MM, Lagnado L (2009) A genetically encoded reporter of synaptic activity in vivo. Nat Methods 6:883-889.

Duemani Reddy G, Kelleher K, Fink R, Saggau P (2008) Three-dimensional random access multiphoton microscopy for functional imaging of neuronal activity. Nat Neurosci 11:713-720.

Durst ME, Zhu G, Xu C (2006) Simultaneous spatial and temporal focusing for axial scanning. Opt Express 14:12243-12254.

Ecker AS, Berens P, Keliris GA, Bethge M, Logothetis NK, Tolias AS (2010) Decorrelated neuronal firing in cortical microcircuits. Science 327:584-587.

Feil R, Brocard J, Mascrez B, LeMeur M, Metzger D, Chambon P (1996) Ligand-activated site-specific recombination in mice. Proc Natl Acad Sci U S A 93:10887-10890.

Ferezou I, Haiss F, Gentet LJ, Aronoff R, Weber B, Petersen CC (2007) Spatiotemporal dynamics of cortical sensorimotor integration in behaving mice. Neuron 56:907-923.

Gandhi SP, Yanagawa Y, Stryker MP (2008) Delayed plasticity of inhibitory neurons in developing visual cortex. Proc Natl Acad Sci U S A 105:16797-16802.

Garaschuk O, Milos RI, Grienberger C, Marandi N, Adelsberger H, Konnerth A (2006) Optical monitoring of brain function in vivo: from neurons to networks. Pflugers Arch 453:385-396.

Göbel W, Helmchen F (2007a) In vivo calcium imaging of neural network function. Physiology (Bethesda, Md) 22:358-365.

Göbel W, Helmchen F (2007b) New angles on neuronal dendrites in vivo. J Neurophysiol 98:3770-3779.

Göbel W, Kampa BM, Helmchen F (2007) Imaging cellular network dynamics in three dimensions using fast 3D laser scanning. Nat Methods 4:73-79.

Gong S, Doughty M, Harbaugh CR, Cummins A, Hatten ME, Heintz N, Gerfen CR (2007) Targeting Cre recombinase to specific neuron populations with bacterial artificial chromosome constructs. J Neurosci 27:9817-9823.

Göppert-Mayer M (1931) Über Elemtarakte mit zwei Quantenspruengen. Ann Phys 9:273.

Greenberg DS, Kerr JN (2009) Automated correction of fast motion artifacts for two-photon imaging of awake animals. J Neurosci Methods 176:1-15.

Grewe BF, Helmchen F (2009) Optical probing of neuronal ensemble activity. Curr Opin Neurobiol 19:520-529.

Grewe BF, Langer D, Kasper H, Kampa BM, Helmchen F (2010) High-speed in vivo calcium imaging reveals neuronal network activity with near-millisecond precision. Nat Methods 7:399-405.

Grinvald A, Hildesheim R (2004) VSDI: a new era in functional imaging of cortical dynamics. Nat Rev Neurosci 5:874-885.

Gutnisky DA, Dragoi V (2008) Adaptive coding of visual information in neural populations. Nature 452:220-224.

Hahnloser RH, Kozhevnikov AA, Fee MS (2002) An ultra-sparse code underlies the generation of neural sequences in a songbird. Nature 419:65-70.

Harris KD, Csicsvari J, Hirase H, Dragoi G, Buzsaki G (2003) Organization of cell assemblies in the hippocampus. Nature 424:552-556.

Hasan MT, Friedrich RW, Euler T, Larkum ME, Giese G, Both M, Duebel J, Waters J, Bujard H, Griesbeck O, Tsien RY, Nagai T, Miyawaki A, Denk W (2004) Functional fluorescent Ca2+ indicator proteins in transgenic mice under TET control. PLoS Biol 2:e163.

Heim N, Griesbeck O (2004) Genetically encoded indicators of cellular calcium dynamics based on troponin C and green fluorescent protein. J Biol Chem 279:14280-14286.

Heim N, Garaschuk O, Friedrich MW, Mank M, Milos RI, Kovalchuk Y, Konnerth A, Griesbeck O (2007) Improved calcium imaging in transgenic mice expressing a troponin C-based biosensor. Nat Methods 4:127-129.

Hell SW (2007) Far-field optical nanoscopy. Science 316:1153-1158.

Helmchen F (2011) Calcium imaging. In: Handbook of Neural Activity Measurement (Brette R, Destexhe A, eds): Cambridge University Press.

Helmchen F, Waters J (2002) Ca2+ imaging in the mammalian brain in vivo. Eur J Pharmacol 447:119-129.

Helmchen F, Denk W (2005) Deep tissue two-photon microscopy. Nat Methods 2:932-940.

Helmchen F, Tank DW (2011) A single compartment model of calcium dynamics in dendrites and nerve terminals. In: Imaging in Neuroscience: A Laboratory Manual (Helmchen F, Konnerth A, eds): Cold Spring Harbor Laboratory Press.

Helmchen F, Imoto K, Sakmann B (1996) Ca2+ buffering and action potential-evoked Ca2+ signaling in dendrites of pyramidal neurons. Biophys J 70:1069-1081.

Hires SA, Tian L, Looger LL (2008) Reporting neural activity with genetically encoded calcium indicators. Brain Cell Biol 36:69-86.

Holekamp TF, Turaga D, Holy TE (2008) Fast three-dimensional fluorescence imaging of activity in neural populations by objective-coupled planar illumination microscopy. Neuron 57:661-672.

Hromadka T, Deweese MR, Zador AM (2008) Sparse representation of sounds in the unanesthetized auditory cortex. PLoS Biol 6:e16.

Hubel DH, Wiesel TN (1959) Receptive fields of single neurones in the cat's striate cortex. J Physiol 148:574-591.

Iyer V, Hoogland TM, Saggau P (2006) Fast functional imaging of single neurons using random-access multiphoton (RAMP) microscopy. J Neurophysiol 95:535-545.

Jia H, Rochefort NL, Chen X, Konnerth A (2010) Dendritic organization of sensory input to cortical neurons in vivo. Nature 464:1307-1312.

Johannssen HC, Helmchen F (2010) In vivo Ca2+ imaging of dorsal horn neuronal populations in mouse spinal cord. J Physiol 588:3397-3402.

Junek S, Chen TW, Alevra M, Schild D (2009) Activity correlation imaging: visualizing function and structure of neuronal populations. Biophys J 96:3801-3809.

Jung R, Creutzfeldt O, Grusser OJ (1957) [Microphysiology of cortical neurons & their role in sensory and cerebral function]. Dtsch Med Wochenschr 82:1050-1059.

Jurrus E, Hardy M, Tasdizen T, Fletcher PT, Koshevoy P, Chien CB, Denk W, Whitaker R (2009) Axon tracking in serial block-face scanning electron microscopy. Med Image Anal 13:180-188.

Kara P, Boyd JD (2009) A micro-architecture for binocular disparity and ocular dominance in visual cortex. Nature 458:627-631.

Kerlin AM, Andermann ML, Berezovskii VK, Reid RC (2010) Broadly tuned response properties of diverse inhibitory neuron subtypes in mouse visual cortex. Neuron 67:858-871.

Kerr JN, Denk W (2008) Imaging in vivo: watching the brain in action. Nat Rev Neurosci 9:195-205.

Kerr JN, Greenberg D, Helmchen F (2005) Imaging input and output of neocortical networks in vivo. Proc Natl Acad Sci U S A 102:14063-14068.

Kerr JN, de Kock CP, Greenberg DS, Bruno RM, Sakmann B, Helmchen F (2007) Spatial organization of neuronal population responses in layer 2/3 of rat barrel cortex. J Neurosci 27:13316-13328.

Kirkby PA, Srinivas Nadella KM, Silver RA (2010) A compact Acousto-Optic Lens for 2D and 3D femtosecond based 2-photon microscopy. Opt Express 18:13721-13745.

Kitamura K, Judkewitz B, Kano M, Denk W, Hausser M (2008) Targeted patch-clamp recordings and single-cell electroporation of unlabeled neurons in vivo. Nat Methods 5:61-67.

Komiyama T, Sato TR, O'Connor DH, Zhang YX, Huber D, Hooks BM, Gabitto M, Svoboda K (2010) Learning-related fine-scale specificity imaged in motor cortex circuits of behaving mice. Nature 464:1182-1186.

Kremer Y, Leger JF, Lapole R, Honnorat N, Candela Y, Dieudonne S, Bourdieu L (2008) A spatio-temporally compensated acousto-optic scanner for two-photon microscopy providing large field of view. Opt Express 16:10066-10076.

Kuhn B, Denk W, Bruno RM (2008) In vivo two-photon voltage-sensitive dye imaging reveals top-down control of cortical layers 1 and 2 during wakefulness. Proc Natl Acad Sci U S A 105:7588-7593.

Kurtz R, Fricke M, Kalb J, Tinnefeld P, Sauer M (2006) Application of multiline two-photon microscopy to functional in vivo imaging. J Neurosci Methods 151:276-286.

Kwan AC, Dietz SB, Webb WW, Harris-Warrick RM (2009) Activity of Hb9 interneurons during fictive locomotion in mouse spinal cord. J Neurosci 29:11601-11613.

Lee D, Port NL, Kruse W, Georgopoulos AP (1998) Variability and correlated noise in the discharge of neurons in motor and parietal areas of the primate cortex. J Neurosci 18:1161-1170.

Leybaert L, de Meyer A, Mabilde C, Sanderson MJ (2005) A simple and practical method to acquire geometrically correct images with resonant scanning-based line scanning in a custom-built video-rate laser scanning microscope. J Microsc 219:133-140.

Li B, Acton ST (2007) Active contour external force using vector field convolution for image segmentation. IEEE Trans Image Process 16:2096-2106.

Li Y, Van Hooser SD, Mazurek M, White LE, Fitzpatrick D (2008) Experience with moving visual stimuli drives the early development of cortical direction selectivity. Nature 456:952-956.

Lillis KP, Eng A, White JA, Mertz J (2008) Two-photon imaging of spatially extended neuronal network dynamics with high temporal resolution. J Neurosci Methods 172:178-184.

Luo L, Callaway EM, Svoboda K (2008) Genetic dissection of neural circuits. Neuron 57:634-660.

Lütcke H, Murayama M, Hahn T, Margolis DJ, Astori S, Zum Alten Borgloh SM, Göbel W, Yang Y, Tang W, Kugler S, Sprengel R, Nagai T, Miyawaki A, Larkum ME, Helmchen F, Hasan MT (2010) Optical recording of neuronal activity with a genetically-encoded calcium indicator in anesthetized and freely moving mice. Front Neural Circuits 4:9.

Mank M, Griesbeck O (2008) Genetically encoded calcium indicators. Chem Rev 108:1550-1564.

Mank M, Santos AF, Direnberger S, Mrsic-Flogel TD, Hofer SB, Stein V, Hendel T, Reiff DF, Levelt C, Borst A, Bonhoeffer T, Hubener M, Griesbeck O (2008) A genetically encoded calcium indicator for chronic in vivo two-photon imaging. Nat Methods 5:805-811.

Mao T, O'Connor DH, Scheuss V, Nakai J, Svoboda K (2008) Characterization and subcellular targeting of GCaMP-type genetically-encoded calcium indicators. PLoS ONE 3:e1796.

Mennerick S, Chisari M, Shu HJ, Taylor A, Vasek M, Eisenman LN, Zorumski CF (2010) Diverse voltage-sensitive dyes modulate GABAA receptor function. J Neurosci 30:2871-2879.

Miri A, Daie K, Burdine RD, Aksay E, Tank DW (2011) Regression-based identification of behavior-encoding neurons during large-scale optical imaging of neural activity at cellular resolution. J Neurophysiol 105:964-980.

Miyawaki A (2003) Fluorescence imaging of physiological activity in complex systems using GFP-based probes. Curr Opin Neurobiol 13:591-596.

Miyawaki A, Llopis J, Heim R, McCaffery JM, Adams JA, Ikura M, Tsien RY (1997) Fluorescent indicators for Ca2+ based on green fluorescent proteins and calmodulin. Nature 388:882-887.

Moreaux L, Laurent G (2008) A simple method to reconstruct firing rates from dendritic calcium signals. Front Neurosci 2:176-185.

Mountcastle VB, Davies PW, Berman AL (1957) Response properties of neurons of cat's somatic sensory cortex to peripheral stimuli. J Neurophysiol 20:374-407.

Mukamel Ea, Nimmerjahn A, Schnitzer MJ (2009) Automated analysis of cellular signals from large-scale calcium imaging data. Neuron 63:747-760.

Nagai T, Yamada S, Tominaga T, Ichikawa M, Miyawaki A (2004) Expanded dynamic range of fluorescent indicators for Ca(2+) by circularly permuted yellow fluorescent proteins. Proc Natl Acad Sci U S A 101:10554-10559.

Nagayama S, Zeng S, Xiong W, Fletcher ML, Masurkar AV, Davis DJ, Pieribone VA, Chen WR (2007) In vivo simultaneous tracing and Ca(2+) imaging of local neuronal circuits. Neuron 53:789-803.

Nagy A (2000) Cre recombinase: the universal reagent for genome tailoring. Genesis 26:99-109.

Nevian T, Helmchen F (2007) Calcium indicator loading of neurons using single-cell electroporation. Pflugers Arch 454:675-688.

Niesner R, Andresen V, Neumann J, Spiecker H, Gunzer M (2007) The power of single and multibeam two-photon microscopy for high-resolution and high-speed deep tissue and intravital imaging. Biophys J 93:2519-2529.

Niessing J, Friedrich RW (2010) Olfactory pattern classification by discrete neuronal network states. Nature 465:47-52.

Nikolenko V, Watson BO, Araya R, Woodruff A, Peterka DS, Yuste R (2008) SLM Microscopy: Scanless Two-Photon Imaging and Photostimulation with Spatial Light Modulators. Front Neural Circuits 2:5.

Nimmerjahn A, Mukamel Ea, Schnitzer MJ (2009) Motor behavior activates Bergmann glial networks. Neuron 62:400-412.

Nimmerjahn A, Kirchhoff F, Kerr JN, Helmchen F (2004) Sulforhodamine 101 as a specific marker of astroglia in the neocortex in vivo. Nat Methods 1:31-37.

O'Donovan MJ, Ho S, Sholomenko G, Yee W (1993) Real-time imaging of neurons retrogradely and anterogradely labelled with calcium-sensitive dyes. J Neurosci Methods 46:91-106.

Ohiorhenuan IE, Mechler F, Purpura KP, Schmid AM, Hu Q, Victor JD (2010) Sparse coding and high-order correlations in fine-scale cortical networks. Nature 466:617-621.

Ohki K, Chung S, Ch'ng YH, Kara P, Reid RC (2005) Functional imaging with cellular resolution reveals precise micro-architecture in visual cortex. Nature 433:597-603.

Ohki K, Chung S, Kara P, Hubener M, Bonhoeffer T, Reid RC (2006) Highly ordered arrangement of single neurons in orientation pinwheels. Nature 442:925-928.

Okun M, Lampl I (2008) Instantaneous correlation of excitation and inhibition during ongoing and sensory-evoked activities. Nat Neurosci 11:535-537.

Olivier N, Mermillod-Blondin A, Arnold CB, Beaurepaire E (2009) Two-photon microscopy with simultaneous standard and extended depth of field using a tunable acoustic gradient-index lens. Opt Lett 34:1684-1686.

Otsu Y, Bormuth V, Wong J, Mathieu B, Dugue GP, Feltz A, Dieudonne S (2008) Optical monitoring of neuronal activity at high frame rate with a digital random-access multiphoton (RAMP) microscope. J Neurosci Methods 173:259-270.

Ozden I, Lee HM, Sullivan MR, Wang SS (2008) Identification and clustering of event patterns from in vivo multiphoton optical recordings of neuronal ensembles. J Neurophysiol 100:495-503.

Petersen CC, Grinvald A, Sakmann B (2003) Spatiotemporal dynamics of sensory responses in layer 2/3 of rat barrel cortex measured in vivo by voltage-sensitive dye imaging combined with whole-cell voltage recordings and neuron reconstructions. J Neurosci 23:1298-1309.

Petreanu L, Huber D, Sobczyk A, Svoboda K (2007) Channelrhodopsin-2-assisted circuit mapping of long-range callosal projections. Nat Neurosci 10:663-668.

Poulet JF, Petersen CC (2008) Internal brain state regulates membrane potential synchrony in barrel cortex of behaving mice. Nature 454:881-885.

Ranganathan GN, Koester HJ (2010) Optical recording of neuronal spiking activity from unbiased populations of neurons with high spike detection efficiency and high temporal precision. J Neurophysiol 104:1812-1824.

Rochefort NL, Garaschuk O, Milos RI, Narushima M, Marandi N, Pichler B, Kovalchuk Y, Konnerth A (2009) Sparsification of neuronal activity in the visual cortex at eye-opening. Proc Natl Acad Sci U S A 106:15049-15054.

Rothschild G, Nelken I, Mizrahi A (2010) Functional organization and population dynamics in the mouse primary auditory cortex. Nat Neurosci 13:353-360.

Saggau P (2006) New methods and uses for fast optical scanning. Curr Opin Neurobiol 16:543-550.

Saito T, Nakatsuji N (2001) Efficient gene transfer into the embryonic mouse brain using in vivo electroporation. Dev Biol 240:237-246.

Sakai R, Repunte-Canonigo V, Raj CD, Knopfel T (2001) Design and characterization of a DNA-encoded, voltage-sensitive fluorescent protein. Eur J Neurosci 13:2314-2318.

Salzberg BM, Grinvald A, Cohen LB, Davila HV, Ross WN (1977) Optical recording of neuronal activity in an invertebrate central nervous system: simultaneous monitoring of several neurons. J Neurophysiol 40:1281-1291.

Sasaki T, Takahashi N, Matsuki N, Ikegaya Y (2008) Fast and accurate detection of action potentials from somatic calcium fluctuations. J Neurophysiol 100:1668-1676.

Sato TR, Gray NW, Mainen ZF, Svoboda K (2007) The functional microarchitecture of the mouse barrel cortex. PLoS Biol 5:e19.

Sawinski J, Wallace DJ, Greenberg DS, Grossmann S, Denk W, Kerr JN (2009) Visually evoked activity in cortical cells imaged in freely moving animals. Proc Natl Acad Sci U S A 106:19557-19562.

Shigetomi E, Kracun S, Sofroniew MV, Khakh BS (2010) A genetically targeted optical sensor to monitor calcium signals in astrocyte processes. Nat Neurosci 13:759-766.

Siegel MS, Isacoff EY (1997) A genetically encoded optical probe of membrane voltage. Neuron 19:735-741.

So PT, Dong CY, Masters BR, Berland KM (2000) Two-photon excitation fluorescence microscopy. Annu Rev Biomed Eng 2:399-429.

Sohya K, Kameyama K, Yanagawa Y, Obata K, Tsumoto T (2007) GABAergic neurons are less selective to stimulus orientation than excitatory neurons in layer II/III of visual cortex, as revealed by in vivo functional Ca2+ imaging in transgenic mice. J Neurosci 27:2145-2149.

Song S, Sjostrom PJ, Reigl M, Nelson S, Chklovskii DB (2005) Highly nonrandom features of synaptic connectivity in local cortical circuits. PLoS Biol 3:e68.

Stettler DD, Axel R (2009) Representations of odor in the piriform cortex. Neuron 63:854-864.

Stosiek C, Garaschuk O, Holthoff K, Konnerth A (2003) In vivo two-photon calcium imaging of neuronal networks. Proc Natl Acad Sci U S A 100:7319-7324.

Sullivan MR, Nimmerjahn A, Sarkisov DV, Helmchen F, Wang SS (2005) In vivo calcium imaging of circuit activity in cerebellar cortex. J Neurophysiol 94:1636-1644.

Svoboda K, Yasuda R (2006) Principles of two-photon excitation microscopy and its applications to neuroscience. Neuron 50:823-839.

Svoboda K, Denk W, Kleinfeld D, Tank DW (1997) In vivo dendritic calcium dynamics in neocortical pyramidal neurons. Nature 385:161-165.

Tamamaki N, Yanagawa Y, Tomioka R, Miyazaki J, Obata K, Kaneko T (2003) Green fluorescent protein expression and colocalization with calretinin, parvalbumin, and somatostatin in the GAD67-GFP knock-in mouse. J Comp Neurol 467:60-79.

Tian L, Hires SA, Mao T, Huber D, Chiappe ME, Chalasani SH, Petreanu L, Akerboom J, McKinney SA, Schreiter ER, Bargmann CI, Jayaraman V, Svoboda K, Looger LL (2009) Imaging neural activity in worms, flies and mice with improved GCaMP calcium indicators. Nat Methods 6:875-881.

Tsien RY (1989) Fluorescent probes of cell signaling. Annu Rev Neurosci 12:227-253.

Turaga SC, Murray JF, Jain V, Roth F, Helmstaedter M, Briggman K, Denk W, Seung HS (2010) Convolutional networks can learn to generate affinity graphs for image segmentation. Neural Comput 22:511-538.

Valmianski I, Shih AY, Driscoll JD, Matthews DW, Freund Y, Kleinfeld D (2010) Automatic identification of fluorescently labeled brain cells for rapid functional imaging. J Neurophysiol 104:1803-1811.

Verma IM, Weitzman MD (2005) Gene therapy: twenty-first century medicine. Annu Rev Biochem 74:711-738.

Vogelstein JT, Packer AM, Machado TA, Sippy T, Babadi B, Yuste R, Paninski L (2010) Fast nonnegative deconvolution for spike train inference from population calcium imaging. J Neurophysiol 104:3691-3704.

Wachowiak M, Cohen LB (2001) Representation of odorants by receptor neuron input to the mouse olfactory bulb. Neuron 32:723-735.

Wallace DJ, Meyer zum Alten Borgloh S, Astori S, Yang Y, Bausen M, Kugler S, Palmer AE, Tsien RY, Sprengel R, Kerr JN, Denk W, Hasan MT (2008) Single-spike detection in vitro and in vivo with a genetic Ca2+ sensor. Nat Methods 5:797-804.

Waters J, Helmchen F (2004) Boosting of action potential backpropagation by neocortical network activity in vivo. J Neurosci 24:11127-11136.

Waters J, Larkum M, Sakmann B, Helmchen F (2003) Supralinear Ca2+ influx into dendritic tufts of layer 2/3 neocortical pyramidal neurons in vitro and in vivo. J Neurosci 23:8558-8567.

Wolfe J, Houweling AR, Brecht M (2010) Sparse and powerful cortical spikes. Curr Opin Neurobiol 20:306-312.

Yaksi E, Friedrich RW (2006) Reconstruction of firing rate changes across neuronal populations by temporally deconvolved Ca2+ imaging. Nat Methods 3:377-383.

Zeng S, Lv X, Zhan C, Chen WR, Xiong W, Jacques SL, Luo Q (2006) Simultaneous compensation for spatial and temporal dispersion of acousto-optical deflectors for two-dimensional scanning with a single prism. Opt Lett 31:1091-1093.

Zipfel WR, Williams RM, Webb WW (2003) Nonlinear magic: multiphoton microscopy in the biosciences. Nat Biotechnol 21:1369-1377.

Zohary E, Shadlen MN, Newsome WT (1994) Correlated neuronal discharge rate and its implications for psychophysical performance. Nature 370:140-143.